# Anneal-free ultra-low loss silicon nitride integrated photonics

Debapam Bose[1], Mark W. Harrington[1], Andrei Isichenko[1], Kaikai Liu[1], Jiawei Wang[1], Zachary L. Newman[2], and Daniel J. Blumenthal[1*]

[1]Department of Electrical and Computer Engineering, University of California Santa Barbara, Santa Barbara, CA 93106, USA
[2]Octave Photonics, Louisville, CO 80027, USA

* Corresponding Author (danb@ucsb.edu)

## ABSTRACT

Heterogeneous and monolithic integration of the versatile low loss silicon nitride platform with low temperature materials such as silicon electronics and photonics, III-V compound semiconductors, lithium niobate, organics, and glasses, has been inhibited by the need for high temperature annealing as well as the need for different process flows for thin and thick waveguides. New techniques are needed to maintain the state-of-the-art losses, nonlinear properties, and CMOS compatible processes while enabling this next generation of 3D silicon nitride integration. We report a significant advance in silicon nitride integrated photonics, demonstrating the lowest losses to date for an anneal-free process at a maximum temperature of 250 C, with the same deuterated silane based fabrication flow, for nitride and oxide, for an order of magnitude range in nitride thickness without requiring stress mitigation or polishing. We report record low losses for anneal-free nitride core and oxide cladding, enabling 1.77 dB/m loss and 14.9 million Q for 80 nm nitride core waveguides, more than half an order magnitude lower loss than previously reported 270 C processes, and 8.66 dB/m loss and 4.03 million Q for 800 nm thick nitride. We demonstrate laser stabilization with over 4 orders of magnitude frequency noise reduction using a thin nitride reference cavity. And using a thick nitride micro-resonator, we demonstrate parametric gain and Optical Parametric Oscillation (OPO) with the lowest reported OPO threshold per unit resonator length for low temperature fabricated nitride, and supercontinuum generation over two octaves. These results represent a significant step towards a uniform ultra-low loss silicon nitride homogeneous and heterogeneous platform for both thin and thick waveguides capable of linear and nonlinear photonic circuits and integration with low temperature materials and processes.

# INTRODUCTION

Ultra-low loss silicon nitride photonic integrated circuits[1] (PICs) have the potential to reduce the size, weight, and cost, and improve the reliability of a wide range of applications spanning the visible to infrared, including quantum computing and sensing[2–5], atomic clocks[6,7], atomic navigation[8], metrology[9], and fiber optic communications[10] as well as enabling new portable applications[11]. In addition to replacing costly systems such as lasers and optical frequency combs that are relegated to bulky table-top systems, there is the potential to improve the performance for precision sciences, such as reducing laser frequency noise which is important for the manipulation and interrogation of atom, ions, and qubits[12,13]. The silicon nitride integration platform has enabled a wide range of waveguide and device designs, from thin nitride waveguides that support ultra-low loss dilute optical modes to thick nitride waveguides that are strongly confining and enhance optical nonlinearities. By varying waveguide parameters, such as nitride core thickness and width, it is possible to design characteristics such as loss, dispersion, nonlinearity, and device footprint[14–16]. Leveraging the properties of both thin weakly confining and thick strongly confining waveguides, this platform enables the designer to realize a wide range of components and functions including ultra-low linewidth lasers[17–21], optical frequency combs[22], optical modulators[23,24], tunable lasers and filters[25,26], and atom and ion beam emitters[2,27–29].

Yet, a major transformation in silicon nitride photonics is needed, where the ultra-low loss and wafer-scale CMOS foundry compatible processes of thin nitride structures and nonlinear properties of thick waveguide structures are maintained while adding the heterogeneous functionality of optical gain, high-speed modulation, electronics, and engineered thermal properties, and at the same time providing a uniform anneal-free waveguide fabrication process for both thin and thick structures. Heterogeneous and monolithic integration of thin and thick nitride photonics with materials that cannot withstand high annealing temperatures is inhibited by incompatibility with the high temperature nitride growth and high-temperature post-oxide cladding annealing process used to achieve today's low losses. Heterogeneous and monolithic integration material platforms of interest include silicon photonic circuits[30], GaAs and InP semiconductor circuits[31,32], and nonlinear materials such as lithium niobate[33] and tantalum pentoxide (tantala)[34] as well as materials for thermal engineering such as quartz substrates[35]. For example, efforts to limit the process temperature to under 400 °C can prevent crystallization in nonlinear tantala waveguides[34], enable processing waveguides directly on silicon electronics, silicon photonic circuits[30,36], thin film lithium niobate[33], and III-V semiconductors[31,32,37]. Further limiting processing temperatures to 250 °C enables a much broader class of heterogeneous and monolithic cointegration with organic electronics[38], polymers like polyimide (Kapton)[39], prepackaged electronics[40], and substrates that are damaged under thermal stress like quartz[35,41].

Therefore, heterogeneous and monolithic integration requires a uniform anneal-free silicon nitride fabrication process that can produce a wide range of nitride core thickness waveguides, of over an order of magnitude range, while maintaining the loss and other planar and high-performance platform properties without additional process complexities such as stress mitigation and chemical mechanical polishing (CMP). State of the art thin (< 100 nm) waveguide silicon nitride photonics are essential to achieve the lowest losses that today reach 0.034 dB/m in the infrared[42,43] and sub-dB/m losses in the visible[44]. These dilute mode ultra-low loss thin waveguides are required for precision applications such as laser frequency stabilization and noise reduction, for example integrated waveguide reference cavities yielding 36 Hz integral linewidth[45] and stimulated Brillouin lasers (SBLs)[21] with sub-100 mHz fundamental linewidth[21]. This level of performance is achieved by reducing overlap of the optical mode with the etched nitride sidewalls and employing Low Pressure Chemical Vapor deposited (LPCVD) silicon nitride waveguides patterned on top of a thermal silicon dioxide lower cladding and a Tetraethyl orthosilicate - plasma enhanced chemical vapor deposited (TEOS-PECVD) upper cladding[42,43]. Yet, these processes require nitride growth temperatures as high as 850 °C[46] and annealing temperatures of 1150 °C[1,42,47]. Recent efforts to reduce the process temperatures of these dilute waveguides employed an unannealed deuterated upper cladding oxide, however, still required 1050 °C annealing of the LPCVD nitride core in order to yield losses of 1 dB/m[48].

For high optical confinement thick nitride devices, the mainstay of nonlinear optical photonics, losses are determined primarily by sidewall scattering and nitride absorption. Thick core nitride waveguide designs utilize strong confinement to achieve efficient optical nonlinearities[15,49–51], achieving losses as low as 0.4 dB/m and resonator Q as high as 67 million[52], requiring anneal temperatures of 1050 °C and structures for stress mitigation as well as CMP. Research to reduce the processing temperature of thick nitride waveguides has focused on deuterated silicon nitride to lower losses in the nitride core only and has not addressed lowering the deuterated oxide cladding losses[53–58].

Therefore, these processes are not capable of realizing ultra-low loss and high resonator Q thin core (< 100 nm) waveguides and devices. Examples of low temperature thick nitride waveguides include 270°C deuterated nitride with losses down to 22 dB/m and quality factors of 1.6 million for partially etched 920 nm thick waveguides[55,59]. More recently, 270 °C deuterated nitride yielded 6 dB/m loss and 5.3 million intrinsic Q in 850 nm thick waveguides for 480 μm radius resonators and 11.9 dB/m loss and 2.9 million intrinsic Q for 150 μm radius resonators (see Supplementary Section S6) [57], and thick waveguide deuterated Si-rich nitride waveguides demonstrated losses of 150 dB/m and resonator intrinsic Q of $1.32 \times 10^5$ with a 350 °C process[53,54,60]. Hydrogen-free low temperature sputtering has also been employed, combined with 300 °C deposited upper cladding, to achieve 32 dB/m losses and 1.1 million intrinsic Q in 750 nm core waveguides[61]. After 400 °C annealing these achieved 5.4 dB/m loss and 6.2 million intrinsic Q. These low temperature processes were used to demonstrate efficient optical nonlinearities including Kerr microcombs[55,56], octave spanning supercontinuum generation[55], and nonlinear frequency generation with Optical Parametric Oscillation (OPO) thresholds of 13.5 mW [56] and OPO threshold per unit resonator lengths down to 23.6 mW/mm [57]. To date, there has not been a demonstration of anneal-free silicon nitride waveguide fabrication, that lowers loss for both the nitride core and oxide cladding, to enable an order of magnitude range of ultra-low loss thin and thick waveguides, with maximum temperature of 250 °C for flexible heterogeneous and monolithic integration.

In this work we report a significant advance in silicon nitride integrated photonics, achieving the lowest loss to date for an anneal-free silicon nitride waveguide. Additionally, using a maximum oxide and nitride temperature of 250 °C we demonstrate the dual use capability for ultra-low loss linear and nonlinear waveguides, using the exact same fabrication process for waveguides with an order of magnitude variation in thickness (80 nm to 800 nm) without any modification in the process flow and without requiring stress mitigation or CMP. We confirm the shifted absorption peaks of our 250 °C grown deuterated $Si_3N_4$ by using Fourier transform infrared (FTIR) spectroscopy (Supplementary Section S2). The 250 °C maximum temperature is compatible with a wide range of materials including organics[38,39]. We report 1.77 dB/m loss and a ~15 million intrinsic Q for thin 80 nm core, over half an order of magnitude lower loss than previous low temperature nitride processes[57,62]. For thick 800 nm waveguides we report comparable to record-low 8.66 dB/m loss and 4.03 million intrinsic Q which is 39 % higher than low temperature deposited thick nitride devices with similar area, as well as resonators that are 7.5 times smaller in area than equivalent record-high Q low temperature fabricated device[57]. To demonstrate the quality of our anneal-free fabrication process, we report record performance linear and nonlinear applications for both ultra-low loss thin and thick nitride waveguides. For thin waveguides we demonstrate a ring resonator optical reference cavity that reduces laser frequency noise by over 4-orders of magnitude using a Pound-Drever-Hall (PDH) lock. We measure 20 $Hz^2$/Hz at a 10 kHz frequency offset from carrier and reduction in the integral linewidth to under 1 kHz, a factor of over 20 times reduction over the free running linewidth. This is the first demonstration of laser stabilization using an anneal-free, low temperature waveguide reference cavity, to the best of our knowledge. This performance is only possible by realizing low loss and high Q, for a 5.36 cm long cavity, almost 20X longer than the longest low temperature processed waveguide reported to date[57], making for a thermorefractive noise (TRN) floor[45] that was $10^3$ times less than that of a typical thick nitride resonator because of the larger modal area of thin waveguides. We also confirm the quality of our 800 nm thick nitride waveguides and resonators with demonstrations of: 1) Resonant optical parametric oscillation (OPO) and Kerr-comb formation and 2) non-resonant supercontinuum generation. Anomalous dispersion is measured, with over 2 octave supercontinuum generation from 650 nm to 2.7 μm as well as four-wave mixing parametric gain with the near-lowest reported threshold of 16.7 mW for silicon nitride waveguides made with a low temperature process. We report an OPO threshold per unit resonator length of 15.2 mW/mm, lower than reported for low temperature deposited thick waveguides [57] and twice as low as deuterated Si-rich thick nitrides (See Supplementary S11 Table TS7)[60,63]. Significantly, our thin waveguide losses are comparable with that of unannealed LPCVD nitride thin core waveguides of the same geometry (Supplementary Section S9). This dual use capability of our anneal-free process, for both thin and thick core linear and nonlinear devices, with high performance loss, demonstrates the versatility of this platform and application to future heterogeneous and monolithic photonic integration.

We illustrate examples of possible heterogeneous and monolithic integration (Fig. 1) enabled by our anneal-free process. These include deposition of ultra-low loss waveguides on III-V semiconductors (Fig. 1a) for high performance lasers and compound semiconductor photonic integrated circuits[64,65], preprocessed electronic circuits and silicon photonics[23,36] (Fig. 1b), organic material based integrated circuits[66] for cointegration with silicon nitride PICs and biophotonics[67] (Fig. 1c), thin film lithium niobate[33] (Fig. 1d), and materials like quartz for athermalization of resonators and reference cavities[35] (Fig. 1e). Additionally, this process can be used to realize sophisticated multi-level silicon nitride photonic circuits[68], homogeneously and monolithically integrated with other materials, to combine high-performance thin-waveguide components like spectrally-pure Brillouin lasers[17] and thick waveguide nonlinear components including optical frequency combs[55,57] (Fig. 1f).

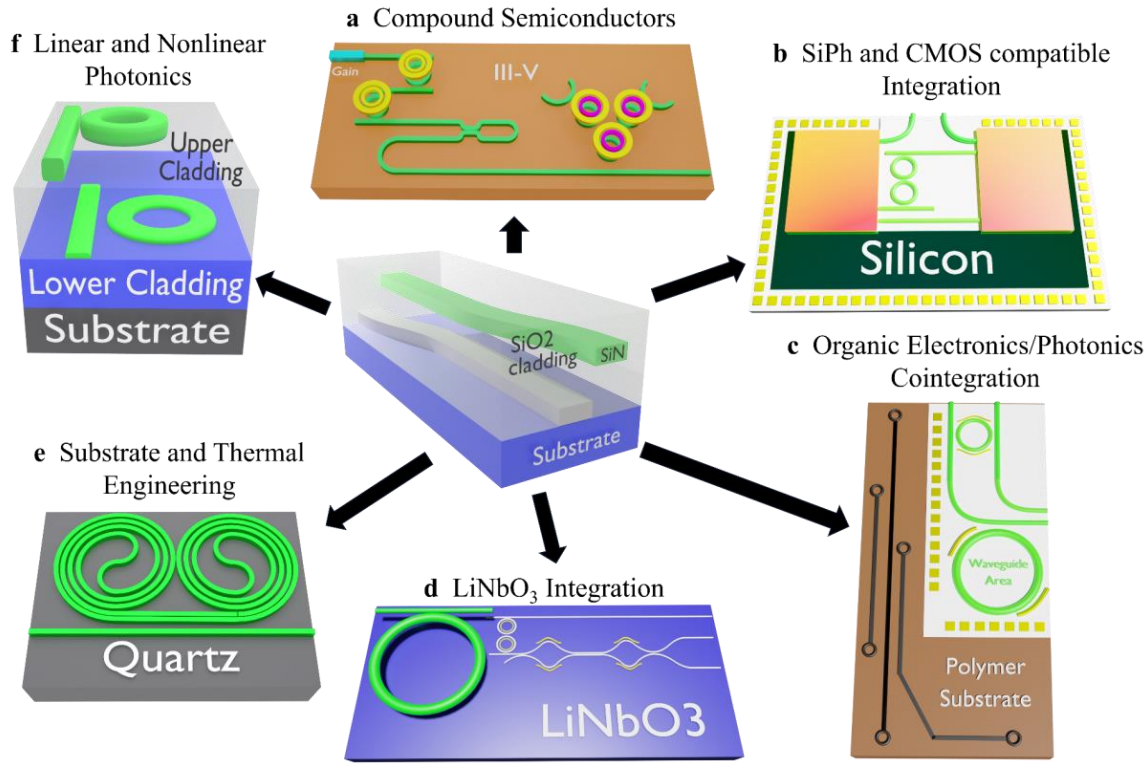

**Figure 1 Examples of different applications of the anneal-free silicon nitride process.** : Cointegration with **a** Compound semiconductors for high performance lasers, **b** preprocessed silicon circuits and silicon photonics, **c** organic electronics/photonics, and **d** thin film lithium niobate. **e** Thermal and substrate engineering such as with quartz substrates. **f** Homogenous integration of thick ( > 650 nm) and thin nitride core devices, each used for different applications.

## RESULTS

### Anneal-Free Fabrication Process and Waveguide Design

In this work we demonstrate that the same process can be used to fabricate oxide clad silicon nitride waveguides that have an order of magnitude variation in nitride core thickness, using the process flow shown in Fig. 2a, realizing the lowest anneal-free silicon nitride waveguide losses to date. This process is used for both thin and thick waveguides, and the wide range of core thickness enables device designs and functions that require different loss regimes and other optical characteristics, such as dispersion, to realize applications as shown in Fig. 2b,c. For example thin ultra-low loss waveguides are required (Fig. 2b) for stimulated Brillouin lasers[17], spiral resonator optical reference cavities[45], and grating beam emitters for creating cold atoms[29,69] and thick nitride waveguides (Fig. 2c) are required for OPO and microcombs[56], supercontinuum generation[55], and mid-IR photonics and gas sensing[70]. The process independence with respect to waveguide thickness, as well as anneal-free maximum temperatures of 250 °C, demonstrates the potential for co-integration of thin to thick nitride core devices and 3D monolithic and homogeneous integration[14,68] as well as monolithic and heterogeneous integration on a variety of other material platforms.

The anneal-free process (Fig. 2a and described in further detail in the Methods Section) starts with a 1 mm thick silicon wafer substrate with pre-processed 15 um thick thermal oxide lower cladding. A uniform silicon nitride layer (e.g. 80 nm or 800 nm) is then deposited using a deuterated silane precursor Inductively Coupled Plasma – Plasma Enhanced Chemical Vapor Deposition (ICP-PECVD) process at 250 °C. The nitride layer is patterned and etched at 50 °C using an Inductively Coupled Plasma Reactive Ion Etcher (ICP-RIE) etch. A final silicon dioxide cladding layer is deposited using the same deuterated silane precursor ICP-PECVD process at 250 °C. In the future, the lower cladding can also be deposited using our 250 °C process for co-integration with other materials and platforms. The different thicknesses of nitride devices correspond to variations in optical mode confinement from dilute modes for thin to strongly confining for thick. The thin waveguide losses are primarily dominated by absorption since nitride sidewall scattering is minimized due to low mode overlap with the core[43], whereas the thick guide losses are dominated by sidewall scattering[14]. The optical mode for thin nitride waveguides exists predominantly in the oxide cladding, therefore it is essential that the anneal-free fabrication process results in low losses for both the deposited nitride and oxide materials.

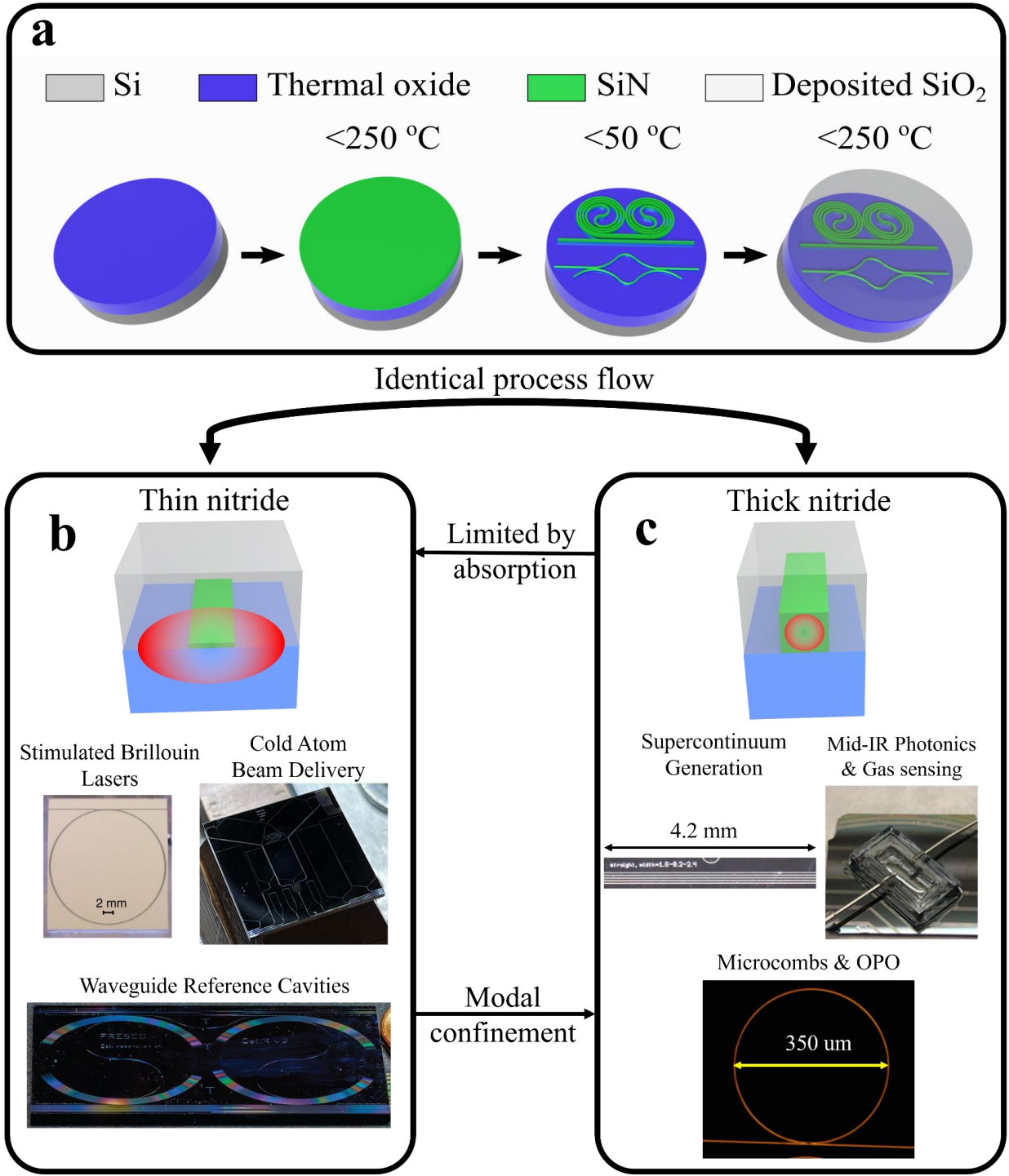

**Figure 2 Anneal-free silicon nitride photonics integration process, thin and thick waveguides, and applications of each. a** Anneal-free fabrication flow. Variation in loss regimes for thin and thick waveguides. **b** Example applications that require the performance of dilute mode thin nitride waveguides, with absorption dominated losses, include stimulated Brillouin lasers[17], coil resonator reference cavities[45], and cold atom trap beam delivery[29,69]. **c** Examples applications that require the characteristics of strongly confining thick nitride waveguides, with higher scattering dominated losses, include supercontinuum generation, mid-IR photonics and gas sensing[70], and microcombs and Optical Parametric Oscillators(OPO).

We characterize the composition of our anneal-free deposited nitride using FTIR[47,56]. We use a silicon nitride bus-coupled ring resonator configuration to access the anneal-free thin nitride losses and compare it to devices made with an unannealed LPCVD silicon nitride process (see Supplementary section S9). The thin nitride waveguide design is a 6 μm wide, 80 nm thick $Si_3N_4$ waveguide core with a 15 μm thick thermal oxide $SiO_2$ lower cladding layer and 5 μm thick oxide upper cladding layer (Fig. 3a) for both the ring and bus waveguides. The ring radius is 8530.8 μm for the thin nitride chip as shown in the example reference resonator photograph in Fig. 3b and the ring-bus coupling gap is 3.45 μm as measured with Scanning Electron Microscopy (SEM) prior to upper cladding deposition (Fig. 3c). The thin core waveguide is designed to support one quasi- Transverse Electric (TE) and one quasi- Transverse Magnetic (TM) mode irrespective of process parameter variations (see Supplementary Section S4). The waveguide design used for our anneal-free process is the same as that used in our standard fully annealed LPCVD nitride and TEOS-PECVD $SiO_2$ process[71]. The thick nitride devices have an 800 nm thick nitride core, a 15 μm thick thermal oxide $SiO_2$ lower cladding layer, and a 4 μm thick oxide upper cladding layer (Fig. 3d). Design splits of the thick nitride devices include ring resonators with waveguide widths varying from 1.4 to 2.4 μm for both ring resonator and bus waveguides, ring radii varying from 165 to 177 μm, and ring-bus coupling gaps varying from 200 to 600 nm. Spiral waveguides were also fabricated with lengths of up to 35 cm, with a sample chip shown in the photograph in Fig. 3e. An example top-down SEM image of a ring resonator with a designed 2 μm waveguide width and 300 nm ring-to-bus waveguide gap is shown in Fig. 3f, indicating high quality thick nitride deposition. Cross-sectional SEM images of the thick nitride core are also provided in Supplementary S2 Fig. S1.

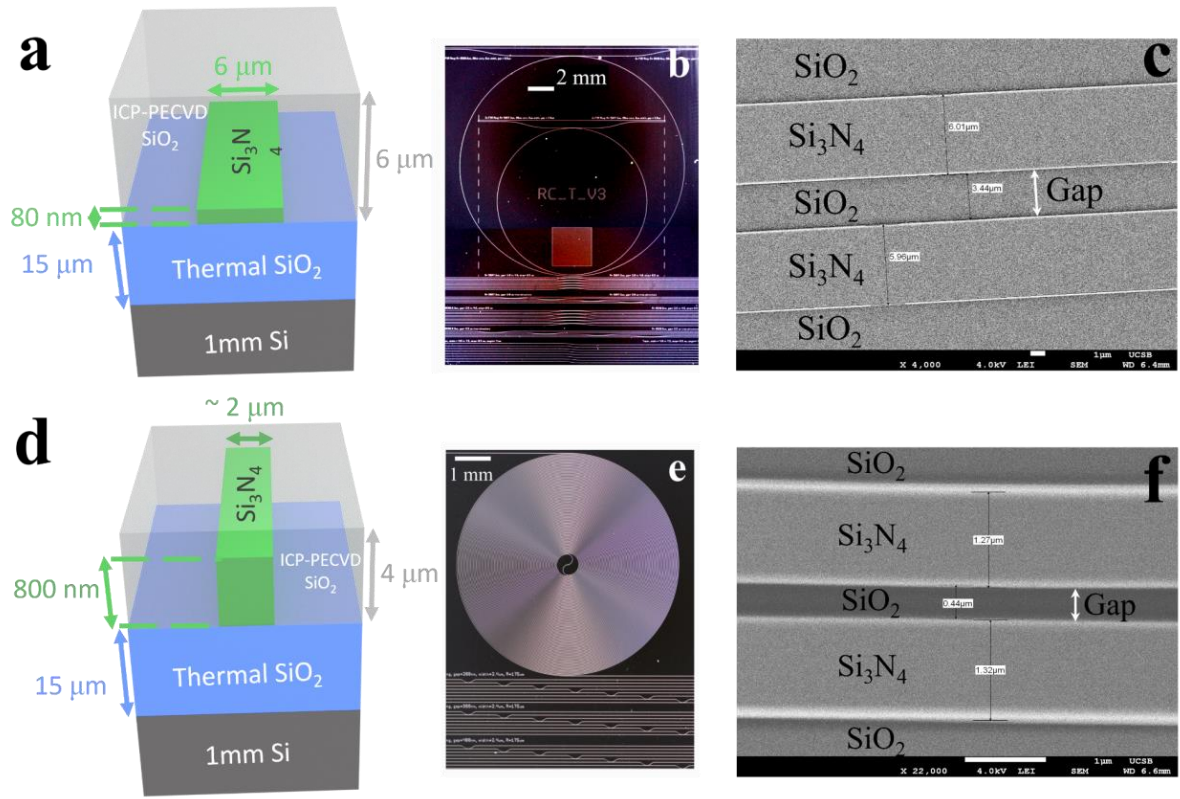

**Figure 3 Material and device characterization and geometries for thin and thick nitride anneal-free devices. a** Thin nitride waveguide geometry. **b** Thin nitride chip showing ring resonators, ring-bus coupling and other test structures. **c** Top-down Scanning Electron Microscopy (SEM) image of the thin nitride waveguide with a width of 6 μm and gap of 3.5 μm on mask. Measured gap is 3.44 μm, and waveguide widths 6.01 and 5.96 μm respectively. **d** Thick nitride waveguide geometry. **e** A thick nitride chip showing a bend loss spiral, ring-bus coupling and other test

structures. **f** Top-down SEM image of a 800 nm thick nitride waveguide ring resonator, with a waveguide width of 1.4 μm and gap of 400 nm on mask. The measured gap is 0.44 μm and measured waveguide widths are 1.32 μm and 1.27 μm respectively. This confirms the high quality of our thick nitride waveguides.

## Thin Nitride Loss/Q and Laser Reference Cavity Application

The waveguide losses and resonator Q are measured and calculated for the fundamental TM mode only, using a calibrated Mach Zehnder interferometer (MZI) technique[17,19,43], and is described in further details in the Methods section. For each thin-nitride resonator design, we characterize 3 different devices (Devices 1-3) and measure TM loss and Q for each device from 1520 to 1630 nm in steps of 10 nm (Fig. 4a). The minimum loss of 1.77 dB/m and maximum intrinsic Q of 14.9 million are measured at 1550 nm. The fabricated devices are over-coupled at wavelengths above 1540 nm, and the maximum Q corresponds to a 4.0 million loaded Q with 49.1 MHz FWHM resonance width (Fig. 4b). The median of the intrinsic Q and loss throughout the above wavelength ranges is 7.77 million and 3.26 dB/m respectively, while the average intrinsic Q and loss are 7.55 million and 4.31 dB/m respectively. Our lowest losses were over half an order of magnitude improvement compared to previous low temperature deuterated devices[57].

Next, we report on a laser stabilization demonstration using PDH locking of a laser to the resonator device 3 at 1550 nm. The laser frequency noise is measured before and after locking using a calibrated MZI frequency noise discriminator[45] (Fig. 4c). The measurement does not use vibration isolation, acoustic shielding, or temperature control of the resonator reference cavity, hence low frequency noise is dominated by environmental effects. The free-running laser $1/\pi$-integral linewidth[72] is 21.3 kHz and the $\beta$-separation linewidth[72] is 195 kHz. After PDH locking to the thin nitride resonator the $1/\pi$-integral linewidth is reduced to 0.976 kHz, a reduction factor of 22, and the $\beta$-separation integral linewidth is reduced to 6.94 kHz, a reduction factor of 28. Under PDH locking, the frequency noise is reduced by over 4 orders of magnitude at 0.67 kHz and 1.4 kHz frequency offsets and as low as 20 $Hz^2$/Hz at 10 kHz frequency offset from the carrier. Further details of the setup and linewidth calculations can be found in Supplementary Section S8.

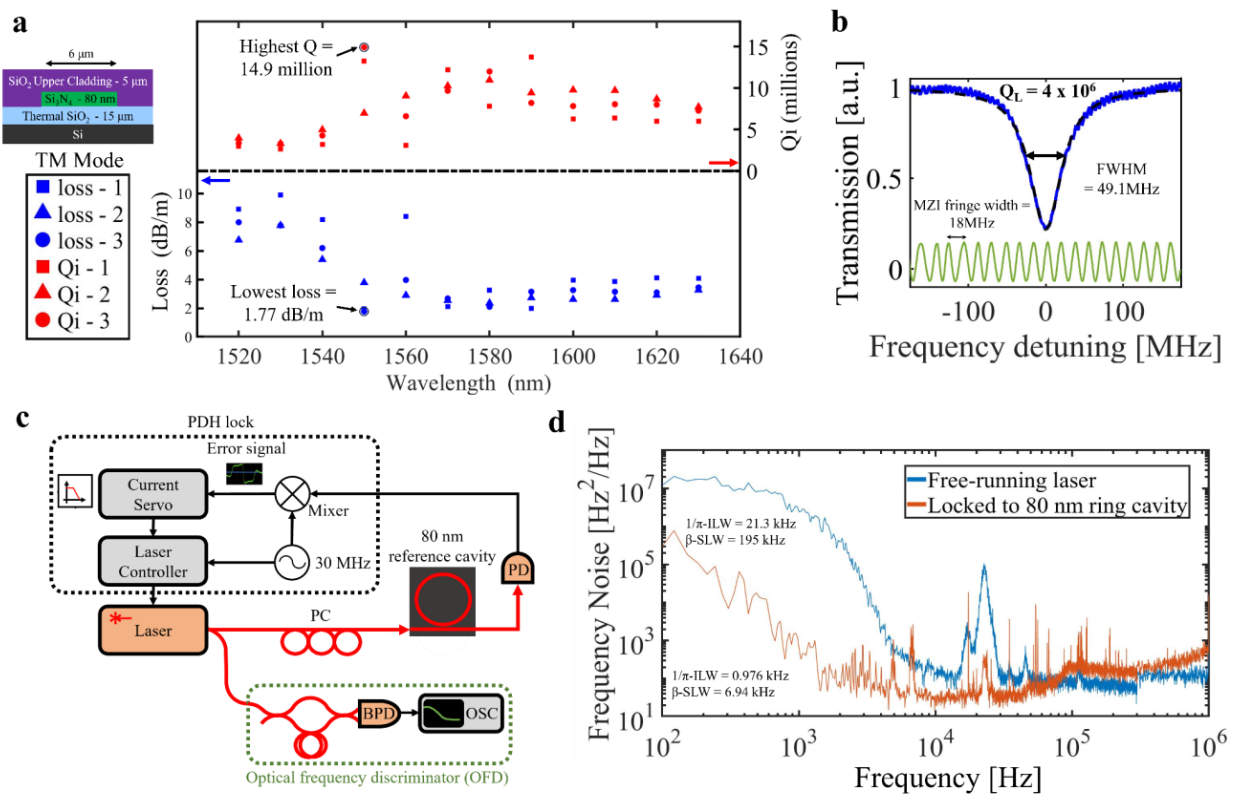

**Figure 4 Thin nitride Q and loss measurements**. **a** The loss and intrinsic Q variation of the Transverse Magnetic (TM) mode vs wavelength for 3 different devices. **b** Q measurement of the TM mode resonance in device 3 at 1550 nm that yields the lowest loss and highest Q of 1.77 dB/m and 14.9 million respectively. The loaded Q and Full Width Max (FWHM) are 4.0 million and 49.1 MHz respectively. **c** The setup for the laser to resonator Pound-Drever-Hall (PDH) locking and frequency noise measurements with an optical frequency discriminator; PD, photodetector; BPD, balanced photodetector; PC, polarization controller; OSC, oscilloscope. **d** Frequency noise measurements at 1550 nm of the laser free-running vs when PDH locked to the ring resonator cavity. The $1/\pi$-integral linewidth ($1/\pi$-ILW) and $\beta$-separation linewidth ($\beta$-SLW) of the laser reduce by factors of 22 and 28 respectively upon locking. The frequency noise was as low as 20 $Hz^2$/Hz at 10 kHz frequency offset from the carrier then.

## Thick Nitride Loss/Q and OPO, FWM, and SHG Nonlinear Photonics Applications

Nonlinear photonic waveguides with wavelength-scale core thickness (~1 μm) offer the high optical confinement and waveguide dispersion needed for effective nonlinearities[15,50,52]. Our 800 nm thick devices are fabricated using exactly the same process flow as described above for the thin nitrides, and the waveguide loss and Q measurements are performed as described in the previous section. We demonstrate that these anneal-free 800 nm nitride waveguides and resonators can achieve: 1) Resonant OPO and Kerr-comb formation and 2) non-resonant supercontinuum generation. We simulate and measure the dispersion (see Supplementary S4 Fig. S3, S11 Fig. S13) and the losses of the different geometry variations in our devices, and based on these results set the waveguide geometry to be 2 μm wide with a 300 nm bus to ring coupling gap, and the resonator radius to 175 μm.

The loss and Q values are measured for a wavelength range of 1550 nm to 1630 nm (Fig. 5a). Example calibrated MZI resonance measurements for the lowest losses are shown in Figs. 5b,c for the TE and TM modes. These measurements yield losses as low as 8.66 dB/m and 16.4 dB/m and intrinsic Q as high as 4.03 million and 2.19 million, for the TE and TM modes respectively. The loaded Q are measured to be 2.30 million and 1.11 with FWHMs of 82.5 MHz and 172 MHz for the TE and TM modes respectively. The median and average intrinsic Q as well as loss for both polarization modes over the measurement wavelength range given in Table TS3 in Supplementary Section S6. Additional Q measurements for the TE mode around the wavelengths where the Q is maximum confirm that the same data points are not due to measurement error (See Supplementary S10 Fig. S10). In fact, these "outlier" wavelengths occur partially because higher order modes do not interact with the fundamental modes[63]. We further perform measurements for the TE mode on 3 different devices with 165 μm radii with the same waveguide width and gap which confirm that the loss and intrinsic Q measurements are repeatable from device to device (See Supplementary S10 Fig. S11).

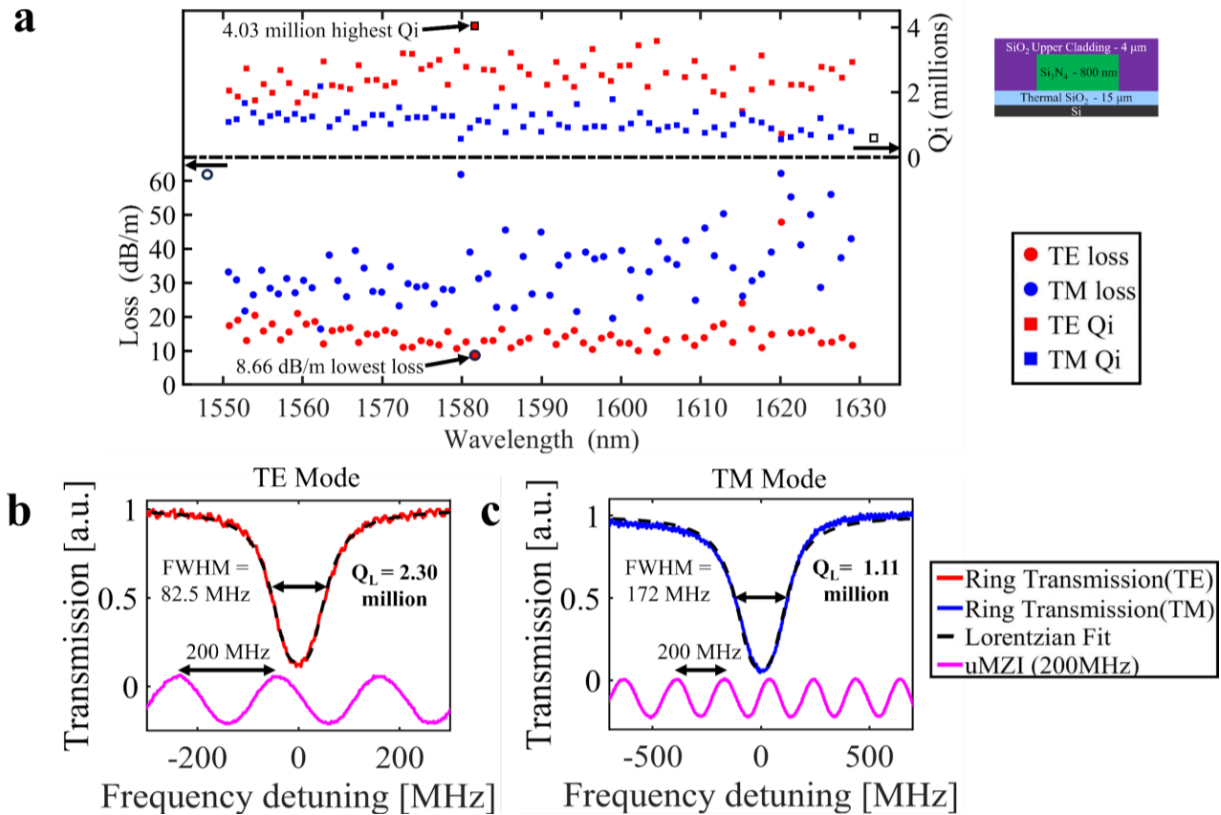

**Figure 5 Thick nitride Q and loss measurements**. **a** Loss and intrinsic Q variation vs wavelength for the Transverse Electric (TE) and Transverse Magnetic (TM) modes from 1550 to 1630 nm wavelengths. **b** The TE mode resonance Q measurement at 1581 nm yielded the lowest loss of 8.66 dB/m and highest intrinsic Q of 4.03 million. The loaded Q and Full Width Max (FWHM) are 2.3 million and 49.1 MHz respectively. **c** The lowest loss TM mode resonance Q measurement at 1560 nm. All measurements shown here are for 175 μm radius ring resonators with 2 μm wide waveguides.

For nonlinear application of the thick nitride we first demonstrate OPO and Kerr-comb formation in a 175 μm radius microring resonator. The resonator has a cross-sectional waveguide of dimensions 800 x 2000 nm (Fig. 6a,b) and we pump a resonance at 1566.7 nm which has a measured $Q_L$ ~ 1.6 million and $Q_i$ ~ 2.0 million. Figure 6b shows an optical micrograph of one device. Fig. 6c shows OPO at an on-chip pump power of 25 mW. As the pump power is increased, Turing pattern formation modulation-instability comb states[22] are also observed (see supplementary section S11). We measure a threshold power, $P_{th}$, for OPO of ~16.7 mW corresponding to an effective nonlinear index, $n_2$ ~ $1.5 \times 10^{-19}$ m$^2$/W (see methods section for more details) which is only slightly lower than typical measurements of $n_2$ for stoichiometric nitride devices[73,74]. This corresponds to the lowest threshold power per unit length of 15.2 mW/mm for any low temperature silicon nitride process (See Table TS7 in Supplementary Section S11).

Next we demonstrate broadband supercontinuum generation in 4 mm long, 800 nm thick straight waveguides (Fig. 6d) with widths ranging from 1.6 to 2.4 um. Fig. 6d shows supercontinuum spectra measured by coupling light from a 1550 nm, 100 MHz repetition rate mode-locked laser with 100 fs pulse duration and on-chip pulse energies ~200-400 pJ into the waveguides. The resulting supercontinuum emission covers two octaves, from ~650 nm to ~2.7 μm. $CO_2$ absorption lines in the spectrum analyzer are evident at the long wave side of the spectrum. While the dispersion of these initial devices is not favorable for mid-infrared supercontinuum generation, we have measured absorption spectra of our deuterated nitride (Supplementary Section S2) and oxide layers[48] and, in principle, our films should support waveguiding and supercontinuum generation out to 4 μm.

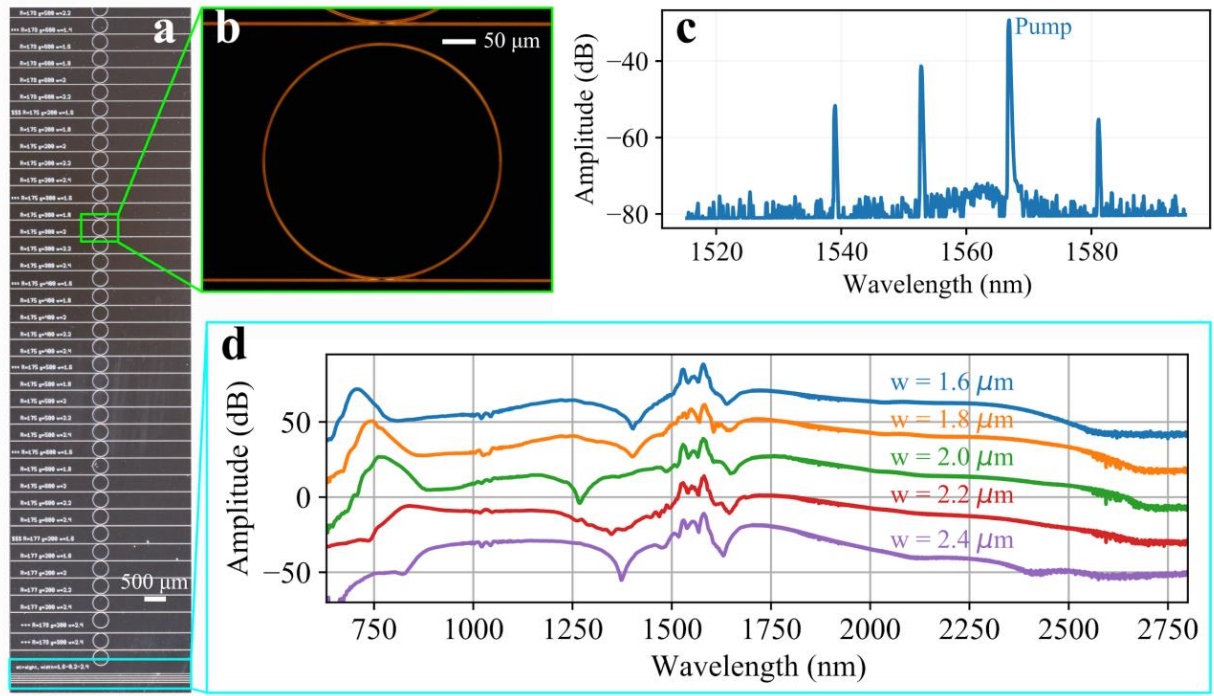

**Figure 6 Nonlinear application demonstrations for anneal-free thick 800 nm waveguides and resonators.** **a** Large field of view image of a thick nitride chip with a broad scan of ring resonator designs and straight waveguides. The green and blue highlighted regions correspond to devices tested in (**c**) and (**d**). **b** Dark-field optical micrograph of the ring resonator device used for Kerr-comb measurement. **c** Optical spectrum of the ring resonator output showing the onset of optical parametric oscillation. **d** Broadband supercontinuum spectra from the different width waveguides highlighted in light blue in (**a**).

# DISCUSSION

We report the lowest loss waveguides and highest Q integrated ring resonators, 1.77 dB/m loss and 15 million Q, fabricated with an anneal-free silicon nitride photonic low temperature process with maximum processing temperature of 250 °C for all steps. We demonstrate that this anneal free process can be used for both thin and thick nitride waveguides, spanning a 10X thickness range, without requiring stress mitigation techniques or chemical mechanical polishing. Using the exact same process as record-low loss thin nitride waveguides, we achieve 8.66 dB/m loss and 4.03 million Q for 800 nm thick nitride waveguides, the highest reported Q for a low temperature processed resonator with equivalent device area (See Supplementary Section S6). We report both linear and nonlinear applications using thin and thick core resonators, demonstrating record performance for both types of applications and an anneal-free fabrication process. Laser noise reduction is demonstrated by PDH locking a laser to an ultra-low loss 80 nm thin nitride resonator employed as an optical reference cavity, achieving 4 orders of magnitude reduction in laser frequency noise. This was possible due to the more than order of magnitude larger modal area of thin nitrides compared to the thick and the long resonator length resulting in a TRN floor which was $10^3$ times smaller than a typical thick nitride resonator (See Supplementary Section S8). A high-Q 800 nm thick nitride resonator is used to achieve resonant OPO with a 16.7 mW threshold corresponding to an OPO threshold per unit resonator length of 15.2 mW/mm and Kerr-comb formation, and over 2 octave non-resonant supercontinuum generation. The low 250 °C temperature and uniformity of this process across waveguide thickness and design, will enable a wide range of systems on-chip applications and novel integration approaches including direct processing on organics, circuit cards, silicon photonic and III-V compound semiconductors, lithium niobate, as well as enabling 3D integration stacking geometries that combine circuits with different nitride core thickness[14,68].

The thin and thick nitride devices cover two different loss regimes, the thin dominated by absorption loss of the cladding material, and the thick by scattering loss and core absorption (Fig. 2b,c). We confirm this for our thin nitride devices at 1550 nm by measuring the thermal bistability for different on-chip powers giving us an absorption loss fraction of 59 % corresponding to an absorption limited loss of 1 dB/m (Supplementary section S7), comparable with 90 nm annealed LPCVD nitride cores with a deuterated oxide cladding[48]. Previously reported work on thick core low temperature nitrides using deuterated processes[55,57,60] as well as sputtering[61] did not demonstrate low absorption loss for their upper claddings and hence ultra-low loss thin nitride devices were not achieved. The absorption losses in our thin nitride devices are thought to be partially from the unannealed thermal oxide lower cladding, which can be further improved by depositing deuterated $SiO_2$ for the lower cladding, a subject of future work. The small amount of hydrogen present in the deuterated silane precursor also increases the absorption loss as evidenced by the increase in waveguide loss towards 1520 nm (Fig. 4a) which is near the 1st overtone of the SiN-H bond absorption. Towards 1630 nm, the loss increase is most likely due to overtones of the SiO-D bond in the upper cladding[48]. We additionally see that the thin losses are comparable to devices of the same geometry made with unannealed LPCVD nitride (Supplementary section S9) confirming that our losses are competitive with respect to process temperature. The more tightly confined modes in the 800 nm thick devices have higher sidewall scattering losses than their thin nitride counterparts, and could be improved by using a hard mask with a smaller grain size such as those made with Atomic Layer Deposition[75,76] or RF sputtering[77]. The TM mode loss for the thick nitride is very different compared to the loss for the TE mode, as the top surface roughness of the nitride core is much lower than the etched sidewall roughness, and the two modes are significantly different in shape (See Supplementary Section S4). It should also be noted that our highest intrinsic Q thick nitride resonances exhibit resonance splitting (see Supplementary Section S6) which is believed to be due to the scattering loss fraction being higher at those resonance wavelengths[43]. In this process we utilized ICP-PECVD with deuterated silane and nitrogen precursors for silicon nitride deposition, avoiding ammonia due to the concentrated Inductively Coupled Plasma (ICP) induced dissociation of $N_2$ that cannot be achieved with conventional parallel plate PECVD [78], and eliminated hydrogen absorption losses[47,55,57,60]. Alternative low temperature processes to ours include sputtering and conventional Plasma Enhanced Chemical Vapor Deposition (PECVD) [79,80] but both suffer from high particle count related scattering losses and conventional PECVD-grown silicon nitride suffers from high hydrogen related absorption losses due to using ammonia and silane precursors[80,81]. Both the linear and nonlinear refractive indices (see Supplementary section S3) are close to that measured for stoichiometric silicon nitride[80].

A summary of published losses and intrinsic Q near the C-band of ring resonators made with different processes as a function of maximum processing temperature and their nitride processing methods is given in Fig. 7, and compared to this work. Our reported lowest losses fall in an "optimum" region between loss and process temperature. It should be noted also that the record low loss thick nitride devices had a width of 10 μm[52]. Our anneal-free process, with a maximum processing temperature of 250 °C, and uniformity for core thickness spanning an order of magnitude, is fully CMOS-compatible and will pave the way to monolithic and heterogeneous integration of ultra-low loss silicon nitride photonics with material systems not possible before such as III-V semiconductors[31,37], lithium niobate[33], preprocessed silicon circuits and photonics[40], and organic electronic materials[39], with applications in metrology[9], navigation[8], telecommunications[10], and quantum information sciences[2–4], and consumer electronics where organic electronics is widely used[82]. This process could also be used to monolithically and homogeneously integrate both thin low confinement and thick high confinement silicon nitride waveguides, enabling 3D integration with optimized device footprint and linear and nonlinear performance. In the future, the temperature of our process has the potential to be modified for as low as 50 °C using further process development on our ICP-PECVD tool[83] (which supports 50 °C processes), enabling the monolithic integration of ultra-low loss photonic integrated circuits on most organic electronic materials.

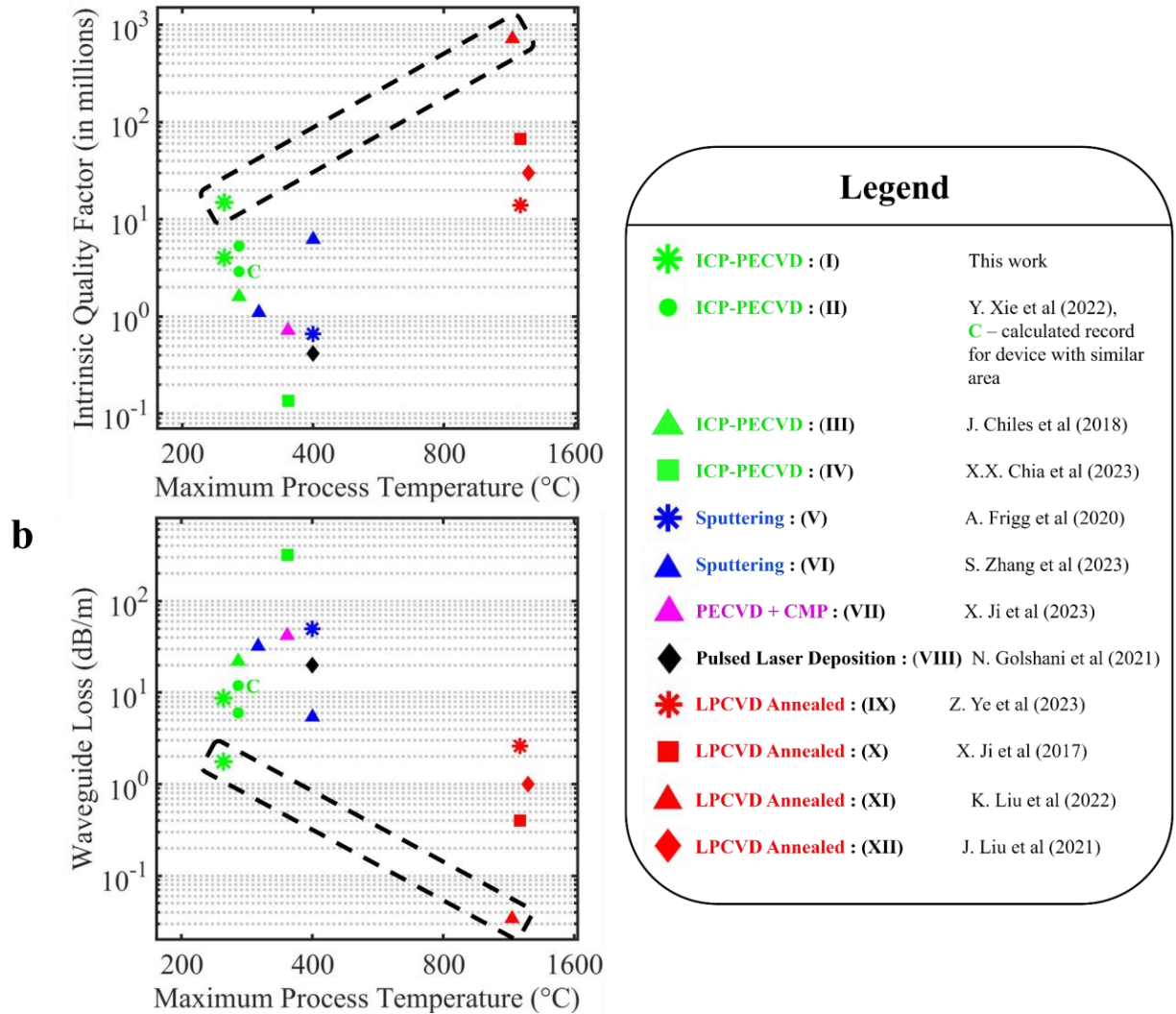

**Figure 7 Q and loss vs temperature near the C-band for different published works based on their silicon nitride growth methods and processing compared to this work.** Our lowest losses (thin nitride) are near an "optimum", denoted by the oval, between low loss and temperature, the current record low loss also being a thin nitride device. Our thick nitride structures have double the Q of the current record calculated for low temperature fabricated devices with similar areas[57] as marked with the C, and very similar loss overall to the absolute record, while having an area 7.5 times smaller. The different works compared include Inductively Coupled Plasma-Plasma Enhanced Chemical Vapor Deposition (ICP-PECVD) processes using deuterated silane precursors like (I) This work, (II) Y. Xie et al.[45], (III) J. Chiles et al.[55], and (IV) X.X. Chia et al.[60] - which also uses Si-rich SiN ; Sputtering such as (V) A. Frigg et al.[79], (VI) S. Zhang et al.[61]; Plasma Enhanced Chemical Vapor Deposition (PECVD) in conjunction with Chemical-Mechanical Polishing(CMP) (VII) X. Ji et al.[84]; Pulsed Laser Deposition - (VIII) N. Golshani et al.[85]; And Low Pressure Chemical Vapor Deposition (LPCVD) together with annealing, such as (IX) Z. Ye et al.[86], (X) X. Ji et al.[52], (XI) K. Liu et al.[42], and (XII) J. Liu et al.[87] - which uses a Damascene process too.

# METHODS

## Fabrication Process

The thick and thin SiN core, and $SiO_2$ upper cladding depositions are performed using an Unaxis VLR ICP-PECVD tool with the same processes used for all core thicknesses and devices. Further details on the nitride deposition and oxide deposition processes can be found in Supplementary section S1. Before any deposition on a device wafer, we run a deposition on a test 100 mm silicon wafer and measure the particle counts, as well as the film thickness and refractive index with an ellipsometer. The deposition on the device wafer is performed only if the particle counts increase by less than 300. The fabrication starts with the 250 °C silicon nitride deposition on Si wafers with 15 μm of thermal oxide, with the thick nitride depositions merely being done for longer than the thin nitride deposition, in a single step. After the nitride deposition step, the thick nitride wafers only get 40 nm of Ruthenium DC sputtered. Both the thick and thin nitride wafers are then patterned in a 248 nm DUV stepper, using the same lithography parameters. The thin nitride is then etched in an ICP-RIE using a $CF_4/CHF_3/O_2$ chemistry, after which it is ashed in a $O_2$ plasma in a ICP tool to remove etch byproducts. Any remaining photoresist is stripped by sonicating in a hot N-methyl-2-pyrrolidone (NMP) solution and rinsing in isopropanol. We additionally perform a standard piranha clean at 100 °C followed by a base piranha (5:1:2 solution of $H_2O:NH_4OH:H_2O_2$) clean at 70 °C , both in freshly prepared solutions, making the thin nitride wafers ready for upper cladding deposition. For the thick nitride fabrication, the Ru on the thick nitride is etched in an ICP-RIE too, to create a hard mask, using a $Cl_2/O_2$ chemistry. The thick nitride wafer is then stripped of photoresist the same way as the thin one, using hot NMP solution and isopropanol. It is then etched in an ICP-RIE using $CF_4$ only, after which the same $O_2$ plasma ashing as the thin nitrides is done. Any remaining Ru is stripped in a wet etch, and then the same piranha cleans done for the thin nitrides are performed. The requisite amount of ICP-PECVD $SiO_2$ upper cladding is then deposited at 250 °C on both the thin and thick nitride wafers. The flow diagrams of these fabrication processes can be found in Supplementary Section S5.

## Quality factor measurements and calculation

The loaded quality factors of the ring resonators are measured using three different calibrated unbalanced fiber MZIs with MZI fringe widths of 5.87 MHz, 18 MHz, and 200 MHz. We have seen in our previous works that Q values measured with this method match well with cavity ring-down measurements[88]. Two Newport Velocity TLB-6700 tunable lasers are used, one with a tuning range of 1520 to 1570 nm, and another one with a tuning range from 1550 to 1630 nm. These lasers are tuned in wavelength with piezo actuators, by applying a ramp signal to the same. A polarization controller is present before the input to the thin nitride devices, which is edge-coupled to a single mode cleaved fiber, while there is a polarization beam splitter present before the input to the thick nitride devices. The full setup for the thin nitride measurements is shown in Supplementary section S6 Fig. S6. Loaded and intrinsic quality factors are extracted by fitting the resonance transmission to a Lorentzian (thin nitride) or coupled-Lorentzian (thick nitride) curves. Coupling and loss parameters are determined by measuring the ring-to-bus couplings on independent ring-bus coupling structures as well as simulating the same[35]. Additional details can be found in supplementary section S6, and plots of all resonance measurements in the Supplementary : Resonance Measurement Summary file.

## Threshold power for optical parametric oscillation

We determine the effective nonlinear index for our deuterated nitride by measuring the threshold power for OPO, $P_{th}$, according to the following[15] :

$$n_2 = \frac{\pi\, n\, \nu_0\, A_{eff}}{8\, P_{th}\, \nu_{FSR}\, Q_i^2} \frac{(1+K)^3}{K}$$

where n is the effective refractive index, $A_{eff}$ is the effective mode area, $\nu_{FSR}$ = 133.5 GHz is the resonator free spectral range, $\nu_o$ is the pump frequency, $Q_i$ is the resonator intrinsic Q, and K is a resonator coupling constant $K = Q_i/Q_c$, where $Q_c$ is the resonator coupling Q. We extract values of $Q_i$ and $Q_c$ through the Lorentzian curve fitting method described above. We then use the software Lumerical MODE to calculate $A_{eff}$ and n as a function of wavelength (in this case 1.35 $\mu m^2$ and 1.85 respectively). Based on our analysis, we determine $n_2 \sim 1.5 \pm 0.2 \times 10^{-19}$ $m^2$/W. Measurement uncertainty is propagated from measurement resolution of the threshold power and the one standard deviation error of the curve fitting parameters which determine Q values.

## DATA AVAILABILITY

The data that support the plots within the paper and other findings of this study are available from the corresponding author upon reasonable request.

## REFERENCES

1. Blumenthal, D. J., Heideman, R., Geuzebroek, D., Leinse, A. & Roeloffzen, C. Silicon Nitride in Silicon Photonics. *Proceedings of the IEEE* **106**, 2209–2231 (2018).

2. Niffenegger, R. J. *et al.* Integrated multi-wavelength control of an ion qubit. *Nature* **586**, 538–542 (2020).

3. Elshaari, A. W., Pernice, W., Srinivasan, K., Benson, O. & Zwiller, V. Hybrid integrated quantum photonic circuits. *Nat. Photonics* **14**, 285–298 (2020).

4. Wang, J., Sciarrino, F., Laing, A. & Thompson, M. G. Integrated photonic quantum technologies. *Nat. Photonics* **14**, 273–284 (2020).

5. Meyer, D. H., Castillo, Z. A., Cox, K. C. & Kunz, P. D. Assessment of Rydberg atoms for wideband electric field sensing. *J. Phys. B: At. Mol. Opt. Phys.* **53**, 034001 (2020).

6. Bloom, B. J. *et al.* An optical lattice clock with accuracy and stability at the $10^{-18}$ level. *Nature* **506**, 71–75 (2014).

7. Newman, Z. L. *et al.* Architecture for the photonic integration of an optical atomic clock. *Optica, OPTICA* **6**, 680–685 (2019).

8. Petrov, A. A. *et al.* Features of magnetic field stabilization in caesium atomic clock for satellite navigation system. *J. Phys.: Conf. Ser.* **1038**, 012032 (2018).

9. Ye, J., Kimble, H. J. & Katori, H. Quantum State Engineering and Precision Metrology Using State-Insensitive Light Traps. *Science* **320**, 1734–1738 (2008).

10. Brodnik, G. M. *et al.* Optically synchronized fibre links using spectrally pure chip-scale lasers. *Nat. Photon.* **15**, 588–593 (2021).

11. Ely, T. A., Burt, E. A., Prestage, J. D., Seubert, J. M. & Tjoelker, R. L. Using the Deep Space Atomic Clock for Navigation and Science. *IEEE Transactions on Ultrasonics, Ferroelectrics, and Frequency Control* **65**, 950–961 (2018).

12. Dick G. J. Local Oscillator Induced Instabilities In Trapped Ion Frequency Standards. in pp 133-147 (Redondo Beach, California, 1987).

13. Audoin, C., Candelier, V. & Diamarcq, N. A limit to the frequency stability of passive frequency standards due to an intermodulation effect. *IEEE Transactions on Instrumentation and Measurement* **40**, 121–125 (1991).

14. Huffman, T. A. Integrated Si3N4 Waveguide Circuits for Single- and Multi-Layer Applications. (University of California, Santa Barbara, Department of Electrical and Computer Engineering, 2018).

15. Briles, T. C., Yu, S.-P., Drake, T. E., Stone, J. R. & Papp, S. B. Generating Octave-Bandwidth Soliton Frequency Combs with Compact Low-Power Semiconductor Lasers. *Phys. Rev. Appl.* **14**, 014006 (2020).

16. Corato-Zanarella, M. *et al.* Widely tunable and narrow-linewidth chip-scale lasers from near-ultraviolet to near-infrared wavelengths. *Nat. Photon.* **17**, 157–164 (2023).

17. Gundavarapu, S. *et al.* Sub-hertz fundamental linewidth photonic integrated Brillouin laser. *Nature Photon* **13**, 60–67 (2019).

18. Jin, W. *et al.* Hertz-linewidth semiconductor lasers using CMOS-ready ultra-high-Q microresonators. *Nat. Photonics* **15**, 346–353 (2021).

19. Chauhan, N. *et al.* Visible light photonic integrated Brillouin laser. *Nat Commun* **12**, 4685 (2021).

20. Isichenko, A., Chauhan, N., Liu, K., Harrington, M. W. & Blumenthal, D. J. Chip-Scale, Sub-Hz Fundamental Sub-kHz Integral Linewidth 780 nm Laser through Self-Injection-Locking a Fabry-Perot laser to an Ultra-High Q Integrated Resonator. Preprint at https://doi.org/10.48550/arXiv.2307.04947 (2023).

21. Liu, K. *et al.* Integrated photonic molecule Brillouin laser with a high-power sub-100-mHz fundamental linewidth. *Opt. Lett., OL* **49**, 45–48 (2024).

22. Kippenberg, T. J., Gaeta, A. L., Lipson, M. & Gorodetsky, M. L. Dissipative Kerr solitons in optical microresonators. *Science* **361**, eaan8083 (2018).

23. Alexander, K. *et al.* Nanophotonic Pockels modulators on a silicon nitride platform. *Nat Commun* **9**,

3444 (2018).

24. Wang, J., Liu, K., Harrington, M. W., Rudy, R. Q. & Blumenthal, D. J. Silicon nitride stress-optic microresonator modulator for optical control applications. *Opt. Express, OE* **30**, 31816–31827 (2022).

25. Alkhazraji, E., Chow, W. W., Grillot, F., Bowers, J. E. & Wan, Y. Linewidth narrowing in self-injection-locked on-chip lasers. *Light Sci Appl* **12**, 162 (2023).

26. Huffman, T. A. *et al.* Integrated Resonators in an Ultralow Loss Si3N4/SiO2 Platform for Multifunction Applications. *IEEE Journal of Selected Topics in Quantum Electronics* **24**, 1–9 (2018).

27. Hummon, M. T. *et al.* Photonic chip for laser stabilization to an atomic vapor with $10^{-11}$ instability. *Optica, OPTICA* **5**, 443–449 (2018).

28. Spektor, G. *et al.* Universal visible emitters in nanoscale integrated photonics. *Optica, OPTICA* **10**, 871–879 (2023).

29. Isichenko, A. *et al.* Photonic integrated beam delivery for a rubidium 3D magneto-optical trap. *Nat Commun* **14**, 3080 (2023).

30. Tran, M. A. *et al.* Ring-Resonator Based Widely-Tunable Narrow-Linewidth Si/InP Integrated Lasers. *IEEE Journal of Selected Topics in Quantum Electronics* **26**, 1–14 (2020).

31. Verrinder, P. A. *et al.* Gallium Arsenide Photonic Integrated Circuit Platform for Tunable Laser Applications. *IEEE Journal of Selected Topics in Quantum Electronics* **28**, 1–9 (2022).

32. Nicholes, S. C. *et al.* An 8 × 8 InP Monolithic Tunable Optical Router (MOTOR) Packet Forwarding Chip. *Journal of Lightwave Technology* **28**, 641–650 (2010).

33. Shams-Ansari, A. *et al.* Reduced material loss in thin-film lithium niobate waveguides. *APL Photonics* **7**, 081301 (2022).

34. Jung, H. *et al.* Tantala Kerr nonlinear integrated photonics. *Optica, OPTICA* **8**, 811–817 (2021).

35. Zhao, Q. *et al.* Low-loss low thermo-optic coefficient Ta2O5 on crystal quartz planar optical waveguides. *APL Photonics* **5**, 116103 (2020).

36. Xiang, C. *et al.* High-Performance Silicon Photonics Using Heterogeneous Integration. *IEEE Journal*

*of Selected Topics in Quantum Electronics* **28**, 1–15 (2022).

37. Wong, M. S., Nakamura, S. & DenBaars, S. P. Review—Progress in High Performance III-Nitride Micro-Light-Emitting Diodes. *ECS J. Solid State Sci. Technol.* **9**, 015012 (2019).

38. Gumyusenge, A. & Mei, J. High Temperature Organic Electronics. *MRS Advances* **5**, 505–513 (2020).

39. DuPont[TM]. Kapton® Summary of Properties. https://www.dupont.com/content/dam/dupont/amer/us/en/ei-transformation/public/documents/en/EI-10142_Kapton-Summary-of-Properties.pdf.

40. Mahajan, R. *et al.* Co-Packaged Photonics For High Performance Computing: Status, Challenges And Opportunities. *Journal of Lightwave Technology* **40**, 379–392 (2022).

41. He, L. *et al.* Broadband athermal waveguides and resonators for datacom and telecom applications. *Photon. Res., PRJ* **6**, 987–990 (2018).

42. Liu, K. *et al.* Ultralow 0.034 dB/m loss wafer-scale integrated photonics realizing 720 million Q and 380 µW threshold Brillouin lasing. *Opt. Lett., OL* **47**, 1855–1858 (2022).

43. Puckett, M. W. *et al.* 422 Million intrinsic quality factor planar integrated all-waveguide resonator with sub-MHz linewidth. *Nat Commun* **12**, 934 (2021).

44. Chauhan, N. *et al.* Ultra-low loss visible light waveguides for integrated atomic, molecular, and quantum photonics. *Opt. Express, OE* **30**, 6960–6969 (2022).

45. Liu, K. *et al.* 36 Hz integral linewidth laser based on a photonic integrated 4.0 m coil resonator. *Optica, OPTICA* **9**, 770–775 (2022).

46. Sharma, N., Hooda, M. & Sharma, S. K. Synthesis and Characterization of LPCVD Polysilicon and Silicon Nitride Thin Films for MEMS Applications. *Journal of Materials* **2014**, e954618 (2014).

47. Osinsky, A. V. *et al.* Optical loss mechanisms in GeSiON planar waveguides. *Appl. Phys. Lett.* **81**, 2002–2004 (2002).

48. Jin, W. *et al.* Deuterated silicon dioxide for heterogeneous integration of ultra-low-loss waveguides. *Opt. Lett., OL* **45**, 3340–3343 (2020).

49. Okawachi, Y., Kim, B. Y., Lipson, M. & Gaeta, A. L. Chip-scale frequency combs for data communications in computing systems. *Optica, OPTICA* **10**, 977–995 (2023).

50. Perez, E. F. *et al.* High-performance Kerr microresonator optical parametric oscillator on a silicon chip. *Nat Commun* **14**, 242 (2023).

51. Yang, K. Y. *et al.* Bridging ultrahigh-Q devices and photonic circuits. *Nature Photon* **12**, 297–302 (2018).

52. Ji, X. *et al.* Ultra-low-loss on-chip resonators with sub-milliwatt parametric oscillation threshold. *Optica, OPTICA* **4**, 619–624 (2017).

53. Chia, X. X. *et al.* Optical characterization of deuterated silicon-rich nitride waveguides. *Sci Rep* **12**, 12697 (2022).

54. Chia, X. X. & Tan, D. T. H. Deuterated SiNx: a low-loss, back-end CMOS-compatible platform for nonlinear integrated optics. *Nanophotonics* **12**, 1613–1631 (2023).

55. Chiles, J. *et al.* Deuterated silicon nitride photonic devices for broadband optical frequency comb generation. *Opt. Lett., OL* **43**, 1527–1530 (2018).

56. Wu, Z. *et al.* Low-noise Kerr frequency comb generation with low temperature deuterated silicon nitride waveguides. *Opt. Express, OE* **29**, 29557–29566 (2021).

57. Xie, Y. *et al.* Soliton frequency comb generation in CMOS-compatible silicon nitride microresonators. *Photon. Res., PRJ* **10**, 1290–1296 (2022).

58. Aihara, T., Hiraki, T., Nishi, H., Tsuchizawa, T. & Matsuo, S. Single soliton generation with deuterated SiN ring resonator fabricated at low temperature. in *Proceedings of the 2022 Conference on Lasers and Electro-Optics Pacific Rim (2022), paper CThP12D_06* CThP12D_06 (Optica Publishing Group, 2022). doi:10.1364/CLEOPR.2022.CThP12D_06.

59. Chiles, J. *et al.* CMOS-compatible, low-loss deuterated silicon nitride photonic devices for optical frequency combs. in *Conference on Lasers and Electro-Optics (2018), paper SF2A.5* SF2A.5 (Optica Publishing Group, 2018). doi:10.1364/CLEO_SI.2018.SF2A.5.

60. Chia, X. X. *et al.* Low-Power Four-Wave Mixing in Deuterated Silicon-Rich Nitride Ring

Resonators. *Journal of Lightwave Technology* **41**, 3115–3130 (2023).

61. Zhang, S. *et al.* Low-Temperature Sputtered Ultralow-Loss Silicon Nitride for Hybrid Photonic Integration. *Laser & Photonics Reviews* 2300642 doi:10.1002/lpor.202300642.

62. Bose, D., Wang, J. & Blumenthal, D. J. 250C Process for < 2dB/m Ultra-Low Loss Silicon Nitride Integrated Photonic Waveguides. in *Conference on Lasers and Electro-Optics (2022), paper SF3O.1* SF3O.1 (Optica Publishing Group, 2022). doi:10.1364/CLEO_SI.2022.SF3O.1.

63. Ye, Z., Fülöp, A., Helgason, Ó. B., Andrekson, P. A. & Torres-Company, V. Low-loss high-Q silicon-rich silicon nitride microresonators for Kerr nonlinear optics. *Opt. Lett., OL* **44**, 3326–3329 (2019).

64. Blumenthal, D. J. *et al.* Integrated Photonics for Low-Power Packet Networking. *IEEE Journal of Selected Topics in Quantum Electronics* **17**, 458–471 (2011).

65. Smit, M. *et al.* An introduction to InP-based generic integration technology. *Semicond. Sci. Technol.* **29**, 083001 (2014).

66. Koos, C. *et al.* Silicon-Organic Hybrid (SOH) and Plasmonic-Organic Hybrid (POH) Integration. *Journal of Lightwave Technology* **34**, 256–268 (2016).

67. Kohler, D. *et al.* Biophotonic sensors with integrated Si3N4-organic hybrid (SiNOH) lasers for point-of-care diagnostics. *Light Sci Appl* **10**, 64 (2021).

68. Moreira, R., Barton, J., Belt, M., Huffman, T. & Blumenthal, D. Optical Interconnect for 3D Integration of Ultra-Low Loss Planar Lightwave Circuits. in *Advanced Photonics 2013 (2013), paper IT2A.4* IT2A.4 (Optica Publishing Group, 2013). doi:10.1364/IPRSN.2013.IT2A.4.

69. Chauhan, N. *et al.* Photonic Integrated Si3N4 Ultra-Large-Area Grating Waveguide MOT Interface for 3D Atomic Clock Laser Cooling. in *2019 Conference on Lasers and Electro-Optics (CLEO)* 1–2 (2019). doi:10.1364/CLEO_SI.2019.STu4O.3.

70. Zhou, J. *et al.* Detection of volatile organic compounds using mid-infrared silicon nitride waveguide sensors. *Sci Rep* **12**, 5572 (2022).

71. Zhao, Q. *et al.* Integrated reference cavity with dual-mode optical thermometry for frequency

correction. *Optica, OPTICA* **8**, 1481–1487 (2021).

72. Domenico, G. D., Schilt, S. & Thomann, P. Simple approach to the relation between laser frequency noise and laser line shape. *Appl. Opt., AO* **49**, 4801–4807 (2010).

73. Gaeta, A. L., Lipson, M. & Kippenberg, T. J. Photonic-chip-based frequency combs. *Nature Photon* **13**, 158–169 (2019).

74. Ikeda, K., Saperstein, R. E., Alic, N. & Fainman, Y. Thermal and Kerr nonlinear properties of plasma-deposited silicon nitride/silicon dioxide waveguides. *Opt. Express, OE* **16**, 12987–12994 (2008).

75. Aaltonen, T., Alén, P., Ritala, M. & Leskelä, M. Ruthenium Thin Films Grown by Atomic Layer Deposition. *Chemical Vapor Deposition* **9**, 45–49 (2003).

76. Mitchell, W. J., Thibeault, B. J., John, D. D. & Reynolds, T. E. Highly selective and vertical etch of silicon dioxide using ruthenium films as an etch mask. *Journal of Vacuum Science & Technology A* **39**, 043204 (2021).

77. Maurya, D. K., Sardarinejad, A. & Alameh, K. Recent Developments in R.F. Magnetron Sputtered Thin Films for pH Sensing Applications—An Overview. *Coatings* **4**, 756–771 (2014).

78. John, D. D. Etchless Core-Definition Process for the Realization of Low Loss Glass Waveguides. (University of California, Santa Barbara, Department of Electrical and Computer Engineering, 2012).

79. Frigg, A. *et al.* Optical frequency comb generation using low stress CMOS compatible reactive sputtered silicon nitride waveguides. in *Integrated Photonics Platforms: Fundamental Research, Manufacturing and Applications* vol. 11364 72–79 (SPIE, 2020).

80. Yang, C. & Pham, J. Characteristic Study of Silicon Nitride Films Deposited by LPCVD and PECVD. *Silicon* **10**, 2561–2567 (2018).

81. Hainberger, R. *et al.* PECVD silicon nitride optical waveguide devices for sensing applications in the visible and <1μm near infrared wavelength region. in *Integrated Optics: Design, Devices, Systems, and Applications V* vol. 11031 40–47 (SPIE, 2019).

82. Ji, D., Li, T., Hu, W. & Fuchs, H. Recent Progress in Aromatic Polyimide Dielectrics for Organic

Electronic Devices and Circuits. *Advanced Materials* **31**, 1806070 (2019).

83. Shao, Z. *et al.* Ultra-low temperature silicon nitride photonic integration platform. *Opt. Express, OE* **24**, 1865–1872 (2016).

84. Ji, X. *et al.* Ultra-Low-Loss Silicon Nitride Photonics Based on Deposited Films Compatible with Foundries. *Laser & Photonics Reviews* **17**, 2200544 (2023).

85. Golshani, N. *et al.* Low-loss, low-temperature PVD SiN waveguides. in *2021 IEEE 17th International Conference on Group IV Photonics (GFP)* 1–2 (2021). doi:10.1109/GFP51802.2021.9673874.

86. Ye, Z. *et al.* Foundry manufacturing of tight-confinement, dispersion-engineered, ultralow-loss silicon nitride photonic integrated circuits. *Photon. Res., PRJ* **11**, 558–568 (2023).

87. Liu, J. *et al.* High-yield, wafer-scale fabrication of ultralow-loss, dispersion-engineered silicon nitride photonic circuits. *Nat Commun* **12**, 2236 (2021).

88. Blumenthal, D. J. Photonic integration for UV to IR applications. *APL Photonics* **5**, 020903 (2020).

# SUPPLEMENTARY

**Table of Contents**

**Section S1 : ICP-PECVD processes and development.**

**Section S2 : Film material Characterization.**

**Section S3 : Refractive indices of materials.**

**Section S4 : Waveguide mode and dispersion simulations.**

**Section S5 : Fabrication process flow.**

**Section S6 : Quality factor measurement and loss extraction/calculation.**

**Section S7 : Absorption loss estimation**

**Section S8 : Thermorefractive Noise (TRN) Floor Estimation, PDH locking and frequency noise measurements**

**Section S9 : Thin nitride loss comparison with LPCVD nitride.**

**Section S10 : Additional Q/loss measurements of thick nitride devices**

**Section S11 : Calculations and additional non-linear application measurements of thick nitride devices**

**S1. ICP-PECVD processes and development**

The 250 °C nitride deposition step uses deuterated silane, nitrogen, and argon respectively. The deuterated silane used is measured to have an isotropic purity of 99%. Before running any actual device wafer, a seasoning process is run with a non-device wafer to coat the chamber. Further, before the nitride deposition step on a device wafer, an Ar preclean is run with said device wafer in the chamber. Particle counts added to wafers after nitride deposition are measured over a 100 mm wafer for sizes between 160 nm to 1.6 μm to be less than 300 consistently using a KLA/Tencor Surfscan. The nitride film etches at a rate of 7.1 nm/min in a Transene UN2817 buffered HF solution, and the deposition rate of the film in the ICP-PECVD tool is measured to be 42 nm/min using an ellipsometer. For a 336 nm nitride film on a 100 mm silicon wafer, the compressive stress is measured to be 666 MPa using a Tencor Flexus FLX-2320 film stress measurement tool.

The 250 °C oxide deposition step is very similar to the work by Jin et al[1] and uses deuterated silane and oxygen respectively. Seasoning and argon preclean steps are also done before oxide deposition on actual wafers, as well as measurement of particle counts and stress etc. The process is regularly characterized by the UCSB cleanroom also.[2]

**S2. Film Material Characterization**

We also take cross-sectional SEM measurements of our 800nm thick nitride waveguides confirming the dimensions and quality of these waveguides as in Fig. S1 below.

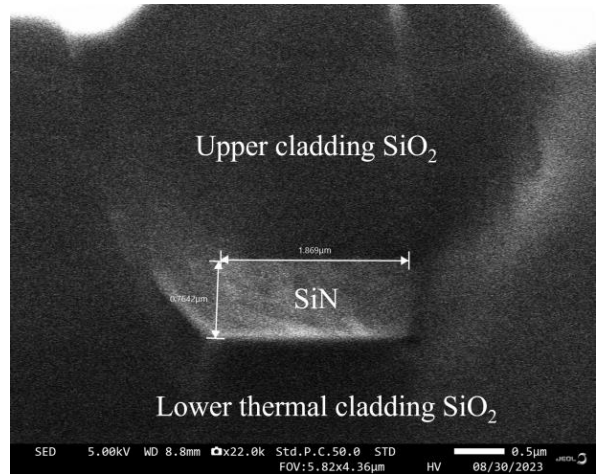

**Figure S1 Cross-sectional SEM of our 800 nm thick nitride waveguides.** Cross-sectional SEM of a 800 nm thick waveguide with 2 μm wide width on mask.

**S3. Refractive indices of materials**

The thin and thick nitride devices are fabricated more than 1.5 years between each other in a university cleanroom, and hence the indices of the deposited materials are slightly different even if using the same recipe, as given below in tables TS1 and TS2. All measurements are from a Woollam Ellipsometer. A refractive index of 1.95 was targeted as that is similar to stoichiometric LPCVD nitride we have used in previous works[5].

**Table TS1. Refractive indices of different materials for thin ICP-PECVD nitride core devices at 1550 nm**

| Material | $Si_3N_4$ | $SiO_2$ lower cladding | Upper cladding ICP-PECVD $SiO_2$ |
|---|---|---|---|
| n | 1.95 | 1.445 | 1.456 |

**Table TS2. Refractive indices of different materials for thick ICP-PECVD nitride core devices at 1550 nm**

| Material | $Si_3N_4$ | $SiO_2$ lower cladding | Upper cladding ICP-PECVD $SiO_2$ |
|---|---|---|---|
| n | 1.963 | 1.445 | 1.459 |

**Section S4 : Waveguide mode and dispersion simulations**

Figure S2 below shows mode simulations for the TM mode for the 80 nm x 6 μm thin and both TE and TM modes for the 800 nm x 2 μm thick nitride core devices respectively, from Lumerical MODE solver.

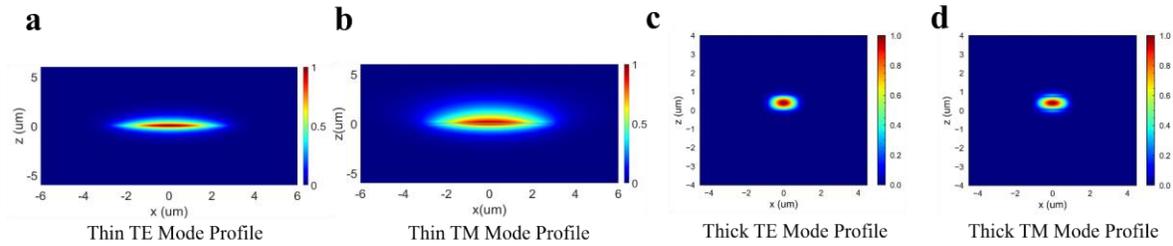

**Figure S2 Mode simulations.** for **a** 80 nm Transverse electric (TE) Mode **b** 80 nm Transverse magnetic (TM) Mode **c** 800 nm TE Mode **d** 800 nm TM Mode. The 800 nm nitride

Below are the effective areas of different modes for the thin and thick nitrides. The order of magnitude larger modal volume of the thin nitride mode makes it ideal for use as a reference cavity, while the smaller thick nitride modes are better for exploiting the Kerr effect.

**Table TS3. Modal area of different modes**

| Mode | TM | TE | TM |
|---|---|---|---|
| **Core Thickness** | 80 nm | 800 nm | 800 nm |
| **Modal Area** | 27 µm² | 1.35 µm² | 1.66 µm² |

Using index data for our nitride and oxide films measured via ellipsometry, Lumerical MODE solver was used to calculate dispersion curves for our 800 nm thick waveguides for a range of widths from 1.5 µm to 2.5 µm, in steps of 0.5 µm for 800 nm of nitride. Figure S3 shows dispersion data for our thick nitride structures which exhibit anomalous dispersion near 1550 nm (red dashed line).

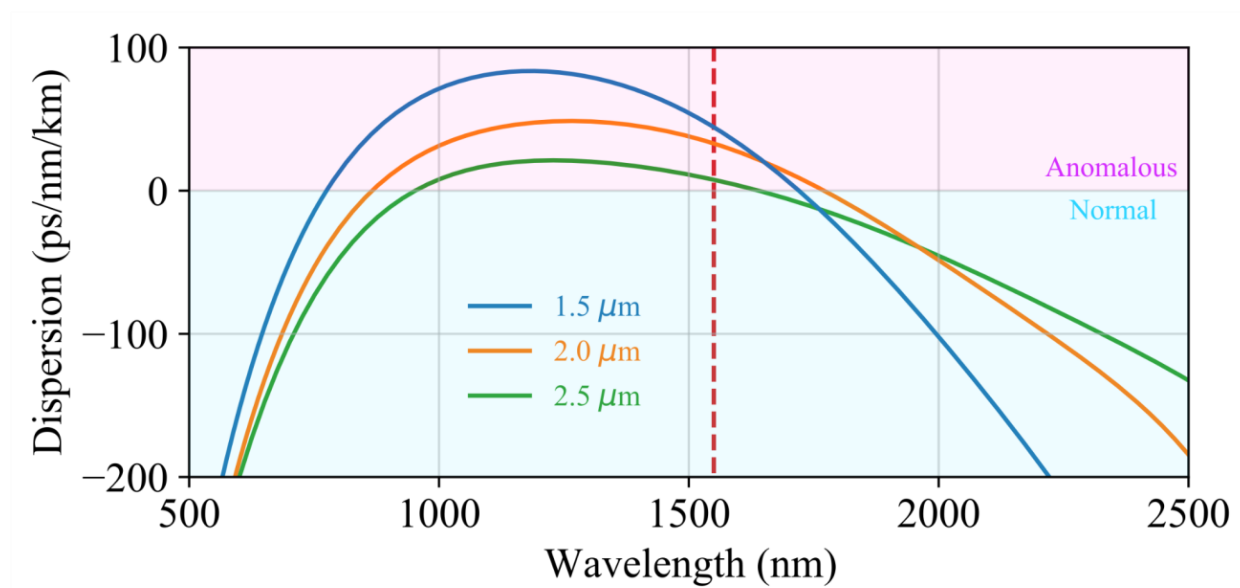

**Figure S3 Waveguide dispersion simulations for different waveguide widths**

### Section S5 : Fabrication process flow

Figure S4 below shows our complete fabrication process flow for our thin nitrides.

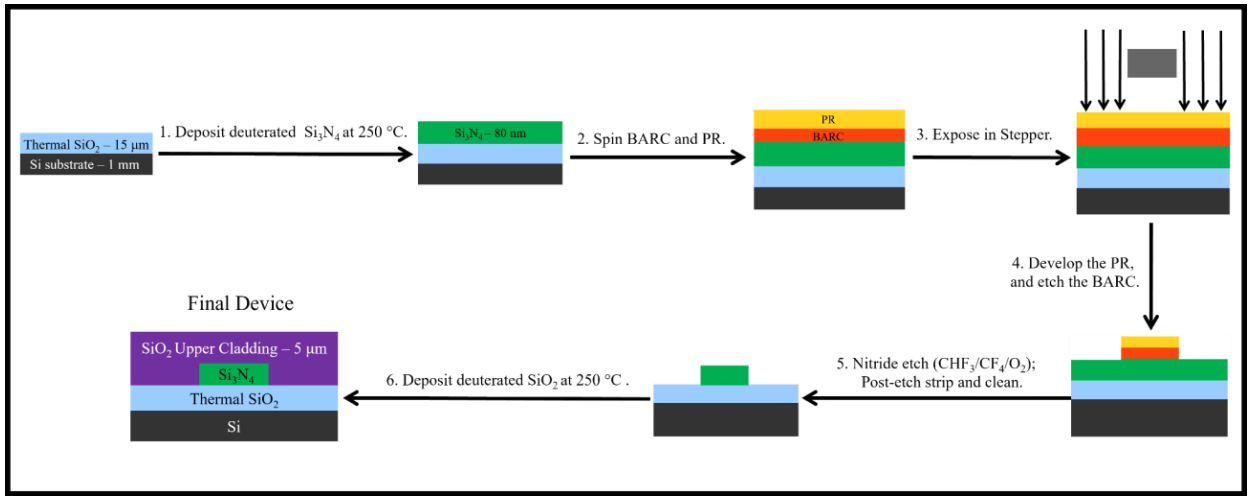

**Figure S4 Detailed fabrication process flow for thin nitride devices.** PR = Photoresist, BARC = Bottom anti-reflective coating.

Figure S5 below shows our complete fabrication process flow for the thick nitrides.

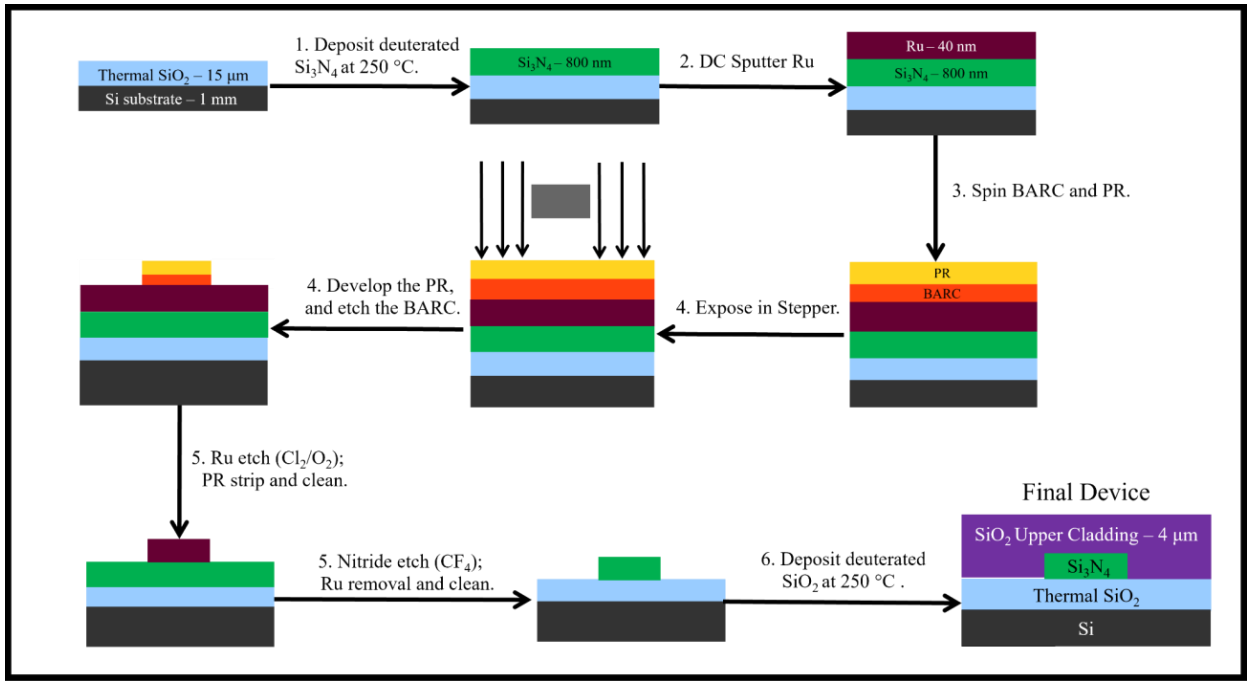

**Figure S5 Detailed fabrication process flow for thick nitride devices**

## S6. Quality factor measurement and loss extraction/calculation

The calibrated Q setup used to measure the Quality factors and loss for the thin nitrides is given Fig. S6 below. The setup for the thick nitrides is exactly the same, except it uses a polarization beam splitter before the Device-Under-Test (DUT).

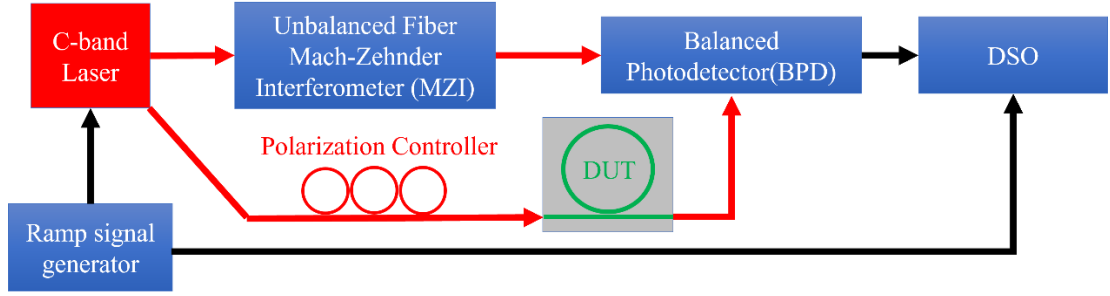

**Figure S6 Calibrated unbalanced Mach-Zehnder Interferometer (MZI) setup for Q factor measurements**

The full-width-at-half-maximum resonance width of the single bus ring resonators is measured with the radio frequency calibrated Mach-Zehnder interferometer (MZI) to extract the quality factor. The propagation loss of the waveguide is extracted based on the following equation[6],

$$Q_{Load} = \frac{\lambda_{res}}{FWHM} = \frac{\pi n_g L \sqrt{ra}}{\lambda_{res}(1-ra)} \quad \text{(ES1)}$$

where $Q_{Load}$ is the loaded quality factor. $n_g$ is the group index of the waveguide, $L = 2\pi R$ is the perimeter of the ring resonator, $\lambda_{res}$ is the resonant wavelength, $r = \sqrt{1-\kappa^2}$ is the self-coupling coefficient and $\kappa^2$ is the power coupling coefficient, $a$ is the single-pass amplitude transmission and is related to the power attenuation coefficient $\alpha$ as $a^2 = exp(-\alpha L)$. The intrinsic Q of the resonator can be calculated with the extraction of waveguide propagation loss $\alpha$ using the following equation[7].

$$Q_{int} = \frac{2\pi n_g}{\lambda_{res} \alpha} \quad \text{(ES4)}$$

The group indexes we use for equation ES4 for loss calculations are from Free Spectral Range (FSR) measurements and are 1.4642 for the 80 nm x 6 μm thin nitride TM mode[8], and 2.025 and 2.053 in the TE and TM modes for the 2 μm wide 800 nm thick devices.

For the ICP-PECVD thin nitride resonators, the TM resonances for all devices below 1550 nm are undercoupled, while almost all resonances 1550 nm and above are overcoupled. As an example to determine whether a TM mode is under or overcoupled, we take the case of our lowest loss 1.77 dB/m resonance at 1550 nm. The simulated ring-bus field coupling (k) using refractive indices from ellipsometry at 1550 nm (Table TS1) and the actual waveguide dimensions (Fig. 3c) using Lumerical FDTD is 0.2141. The undercoupled solution for this resonance gives a k of 0.14, while the overcoupled solution gives a k of 0.2379 which is within the tolerance of our measurement to the simulated value. The overcoupled value of k here also agrees better with measurements of k from ring-bus coupling structures present on the same chip (Fig. 2b). The TE mode resonances for the thin nitrides are all undercoupled, and are difficult to measure accurately to calculate Qi and loss for all wavelengths for all of the devices because of the low extinction of the resonances. For the unannealed Low Pressure Chemical Vapor Deposited (LPCVD) devices in Section S7, the TM modes are all overcoupled.

The resonances for the thick nitride resonators shown are all undercoupled for both the TE and TM modes. These resonances are fit to a modified lorentzian curve to account for resonance splitting caused by backscattering in the ring[9]. Some statistics about the Qs and losses measured of the 2 μm waveguide width and 300 nm gap thick nitride device is given in Table TS3 below.

**Table TS4. Thick nitride median and average intrinsic Q and losses**

| Mode | Median of intrinsic Qs (millions) | Average of intrinsic Qs (millions) | Median of losses (dB/m) | Average of losses (dB/m) |
|---|---|---|---|---|
| TE | 2.59 | 2.60 | 13.9 | 14.8 |
| TM | 1.07 | 1.11 | 32.9 | 34.1 |

The loss and Q of the low temperature thick nitride devices with similar area were calculated using the same split resonance model as we used for our own devices and fitting to the data in the "Fig. 3c" in Y. Xie et al[10]. This resonance had a loaded Q of 1.5 million, at 1560.39 nm, with a Free Spectral Range (FSR) of 150 GHz as given in Y. Xie et al[10], yielding an intrinsic Q of 2.9 million and loss of 11.9 dB/m.

**S7. Absorption loss estimation**

We measure the absorption loss at 1550 nm of thin nitride device 3 to separate the contributions of absorption and scattering losses to the total loss. This measurement follows a technique to quantify the photothermal induced bistable linewidth shift of the longitudinal ring resonances[9], using a spectral scan across the resonance with a high on-chip power, to induce a photothermal resonance redshift that is comparable to the resonance linewidth. This photothermal effect is due to absorption heating in the resonator. As shown in Fig. S7a, the red detuning (from shorter to longer wavelengths) across resonance heats up the resonator, and induces a resonance redshift, resulting in a skewed lineshape. To extract the absorption loss relative to the total loss, we simulate the thermal impedance $R_{th}$ of the ring resonator in Comsol®, giving us $R_{th}$ = 6.87 K/W. We then measure the thermal-optic redshift with a global heating $\Delta f_{res}/\Delta T$ = 1.23 GHz/K which yields the resonance redshift per milliwatt of optical power absorbed by the resonator $\Delta f_{res}/P_{abs}$ = 8.45 MHz/mW. The resonance redshift has a linear relationship with on-chip power (Fig S8b), confirming the photothermal heating effect, from which we extract the absorption loss fraction to be 59 %. This yields an absorption loss of 1.04 dB/m, which can be said to be an upper bound for the absorption loss of the cladding in this device. Some of this absorption loss might come from the 1520 nm loss peak (See W.Jin et al[1] and our(Fig. 4a)) due to SiN-H bonds in the nitride due to residual hydrogen, even for thin nitrides.

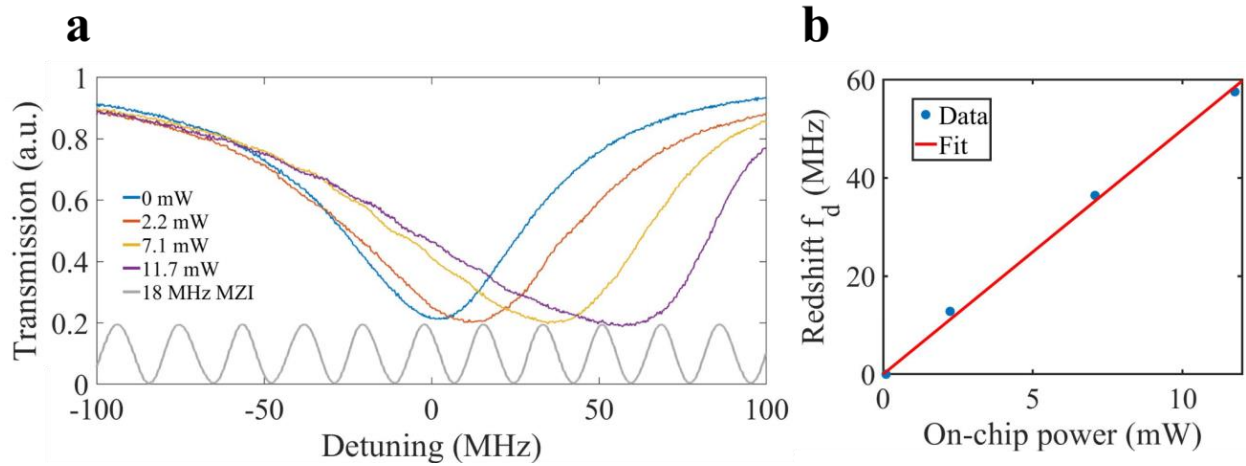

**Figure S7. Photo-thermal heating and absorption loss measurement. a** Photothermal effect is amplified by higher on-chip power and the resonance redshift exhibits a linear relationship with the on-chip power. **b** Normal Lorentzian fitting for the lower power spectral sweeping and skewed Lorentzian fitting for the high power spectral sweeping extracts the intrinsic loss and absorption loss rates.

## S8. Thermorefractive Noise (TRN) Floor Estimation, PDH locking and frequency noise measurements

From the modal area (Table TS3) and cavity length of a device, one can estimate the thermorefractive noise (TRN) floor of said cavity[11]. Our thin nitride cavity is hence estimated to have a TRN floor of around 10 Hz$^2$/Hz which is more than 3 orders of magnitude lower than the TRN floor for a typical thick nitride ring resonator enabling us to lower the frequency noise of a locked laser to a cavity by 3 orders of magnitude more than would be possible with a typical thick nitride device.

**Table TS5. TRN Floor Estimation for different device designs**

| Device | Cavity Length | Mode Area | TRN Floor at $10^4$ Hz offset |
|---|---|---|---|
| K. Liu et al[11] | 4 m | 18 um$^2$ | 0.1 Hz$^2$/Hz |
| Typical thick nitride | 1 mm | 1 um$^2$ | 14,250 Hz$^2$/Hz |
| Our thin cavity | 53.6 mm | 27 um$^2$ | 9.85 Hz$^2$/Hz |

In the case where one makes a thick nitride resonator with as long a length as the thin nitride mode size is larger (27 times), one will still not be able to reach as low frequency noise when locked as the same also depends inversely on the quality factor[11]. Further, increasing a thick nitride resonator length to 1.5 m would increase the waveguide loss typically due to increase in the total number of accumulated defects in the resonator waveguide, further increasing the lowest frequency noise possible.

Laser frequency stabilization was achieved using a standard Pound Drever Hall locking arrangement, as shown in Fig. 4c, and described in detail in K. Liu et al[11]. A velocity TLB 6730-P tunable External Cavity Diode Laser (ECDL) was used as the laser source, with 30 MHz phase modulated sidebands applied via current modulation. Approximately 1 mW optical power was delivered to the resonator, and the resulting transmission signal was photodetected with a 10 kV/A Trans-Inductance Amplifier (TIA) amplified photodetector.

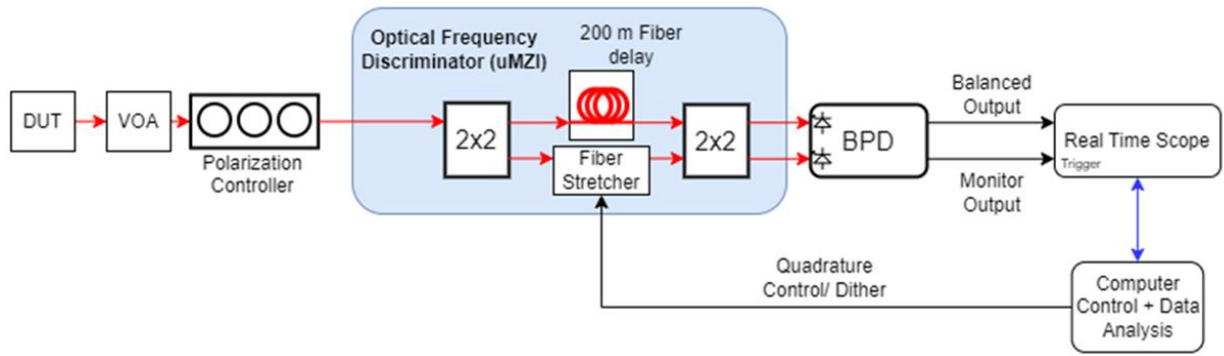

**Figure S8 Detailed system diagram for laser frequency noise measurement using an optical frequency discriminator.**

Frequency noise measurements were carried out using an unbalanced Mach-Zender interferometer with a 200m delay length as in Fig. S8, and described in detail in G. Brodnik et al[12].

The low frequency cutoff for the linewidth calculation was 300 Hz, and the high frequency cutoff was 990,000 Hz.

**S9. Thin nitride loss comparison with LPCVD nitride**

We compare the losses of devices made using the thin nitride geometry (80 nm x 6 μm) between those using deuterated ICP-PECVD nitride cores to those using unannealed Low Pressure Chemical Vapor Deposited (LPCVD) cores, both using the same deuterated upper cladding, for a fair comparison. We see that at 1550 nm and above, the losses are very much comparable.

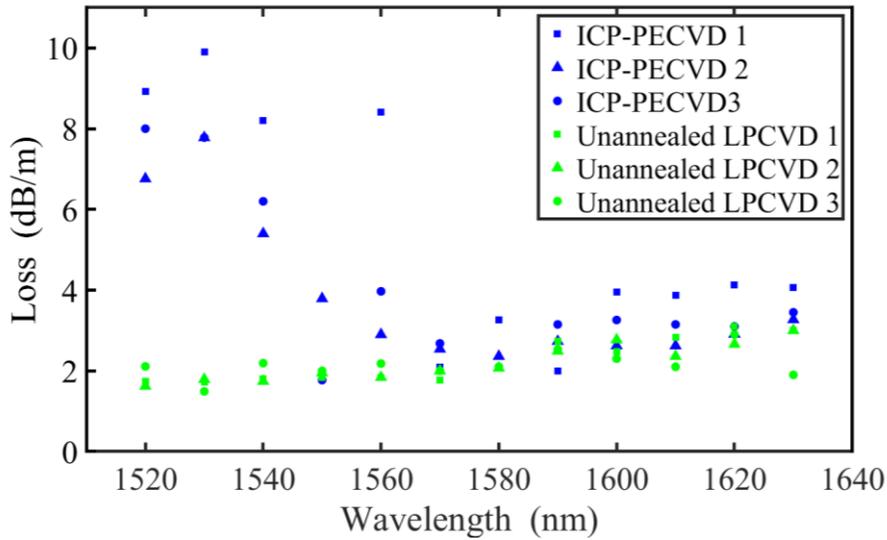

**Figure S9 Loss comparison between thin nitride devices made using different processes.** The losses of our 250 °C Inductively Coupled Plasma-Plasma Enhanced Chemical Vapor Deposition (ICP-PECVD) process devices are similar to those made using unannealed Low Pressure Chemical Vapor Deposition (LPCVD) nitride cores, both using the same deuterated upper cladding.

### S10. Additional Q/loss measurements of thick nitride devices

Additional Q measurements of the thick nitride TE modes for the 175 μm radius device around the wavelengths where the loss minima or outliers in wavelength occur in Fig. 5a are taken (Fig. S10) namely around 1580, 1600, and 1606 nm respectively, showing that these are not measurement discrepancies and may be related to avoided mode crossings with higher order modes.

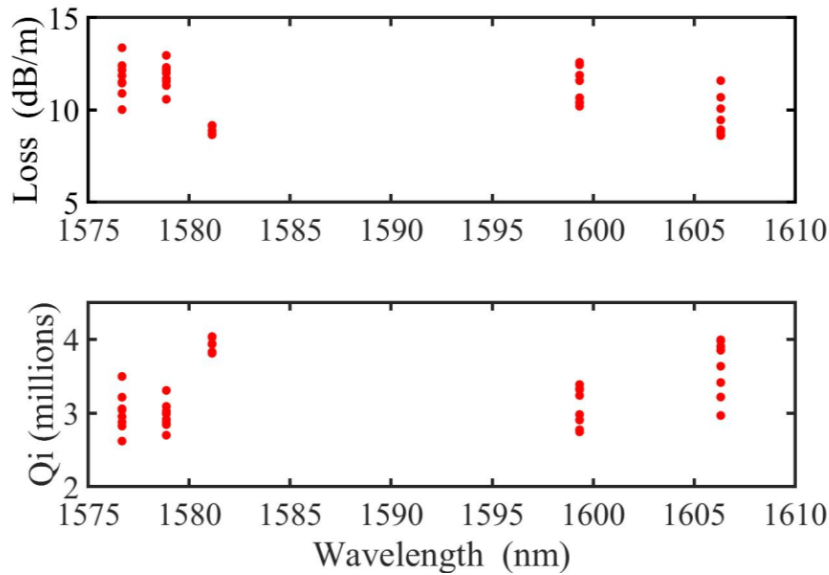

**Figure S10 Additional Q/loss measurements near loss minima.** Additional Q/loss measurements were taken around 1580, 1600, and 1606 nm respectively showing consistency of the Q/loss values between measurements.

We also take more Q/loss measurements of 3 different ring resonator devices with 165 μm radii, but also having a 300 nm ring-bus gap and 2 μm wide waveguide as for the devices in Fig. 5, for the TE modes. The losses and Qs being fairly consistent across the different devices shows that our process is reliable (Fig. S11).

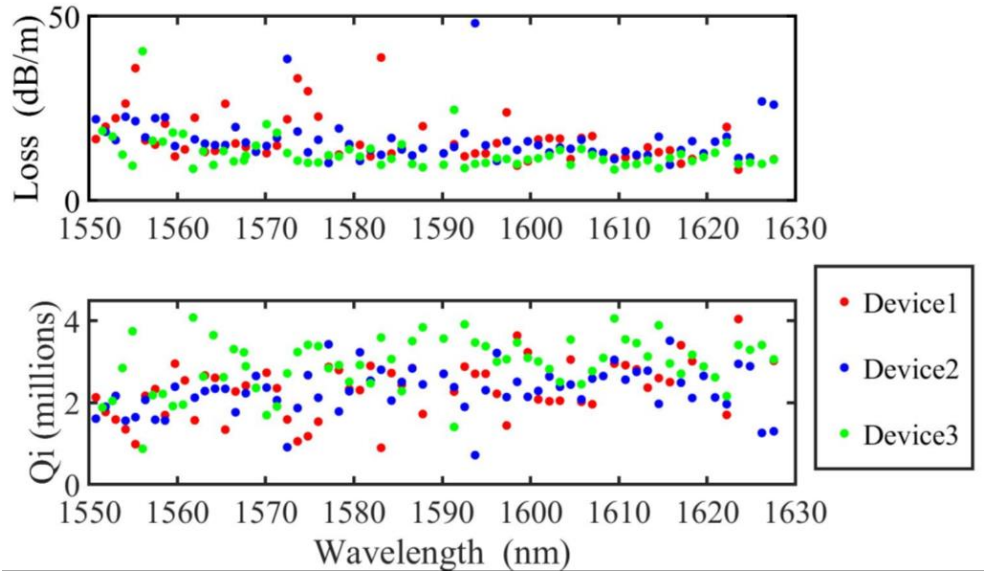

**Figure S11 Loss and Q comparison between 3 different thick nitride ring resonator devices of radius 165 um, 2 um wide waveguide, with 300 nm ring-bus gaps for the TE modes.** Our loss and intrinsic Qs are fairly consistent across the different devices

The median and average losses for TE modes for these devices is given in Table TS5 below.

**Table TS6. Median and average intrinsic Q and losses for the three 165 μm ring resonator devices**

| Mode | Median of intrinsic Qs (millions) | Average of intrinsic Qs (millions) | Median of losses (dB/m) | Average of losses (dB/m) |
|---|---|---|---|---|
| TE | 2.51 | 2.50 | 13.7 | 19.2 |

**S11. Calculations and additional non-linear application measurements of thick nitride devices**

The Optical Parametric Oscillation (OPO) thresholds and thresholds per unit length of various works are calculated and shown below in Table TS4.

**Table TS7. OPO Threshold per unit length of different works**

| OPO threshold (mW) | Ring resonator radius (μm) | OPO threshold per unit length (mW/mm) | Work |
|---|---|---|---|
| 16.7 | 175 | 15.2 | This work. |
| 23.7 | 160 | 23.6 | Y. Xie et al[10] |
| 13.5 | 80 | 26.9 | Z. Wu et al[3] |

| 40 | 23 | 276.8 | J. Chiles et al[13] |
| 10 | 50 | 31.8 | X.X. Chia et al[14] |
| 21 | 100 | 33.42 | Z. Ye et al[15] |

Kerr comb formation was measured using a widely tunable ECDL amplified by a high power EDFA. The laser frequency was tuned to be slightly blue detuned from a TE mode resonance located at 1566.7 nm (at low optical power), and the resonator output was monitored with an optical spectrum analyzer. As optical power was increased, the laser frequency was slowly tuned to maintain the smallest possible blue detuning between laser and resonator. On-chip power was calculated by subtracting half the total throughput coupling loss of the resonator from the measured input optical power to the chip. At on-chip powers higher than 40 mW, The comb transitions into the modulation instability regime, as seen in Figure S12 below. Due to difficulties with strong thermal shifting and a large number of avoided mode crossings, soliton steps were not able to be observed.

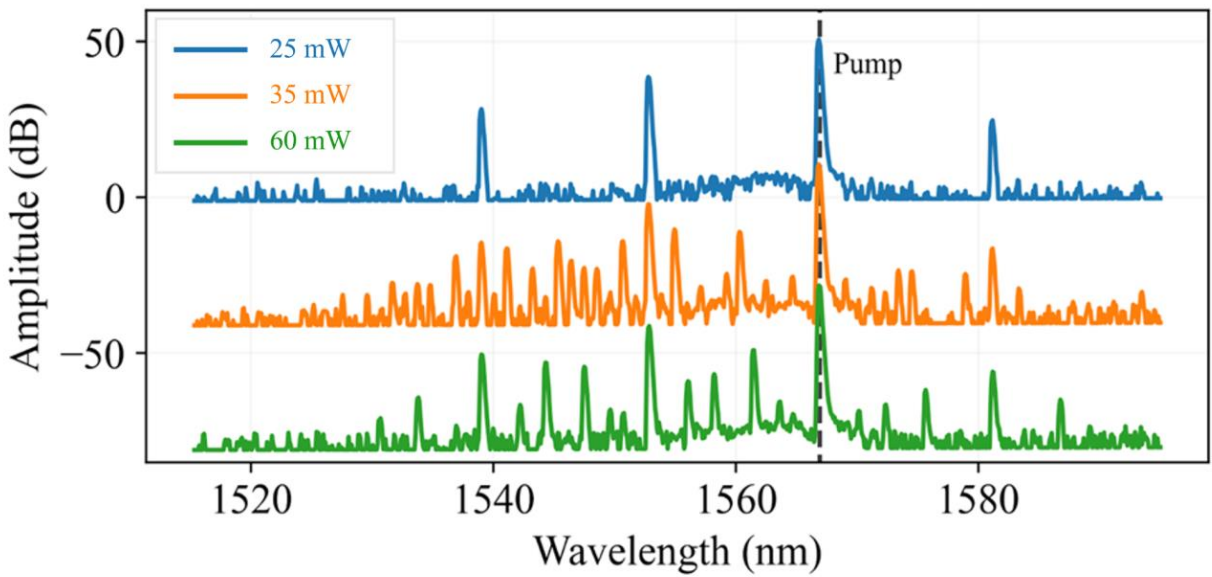

**Figure S12 Kerr comb formation.** Thick nitride Kerr comb evolution at various on-chip pump powers.

To confirm that the device is able to support soliton formation, a temporal lugiato-lefever equation simulation[16] was used to simulate soliton formation dynamics of the 2000 nm TE mode ring-resonator, using dispersion information measured(Fig. S13 (b)) via a widely tunable continuous laser sweep. Figure S13 (a) below shows a laser detuning sweep with 600 mW on-chip power that results in multiple soliton steps.

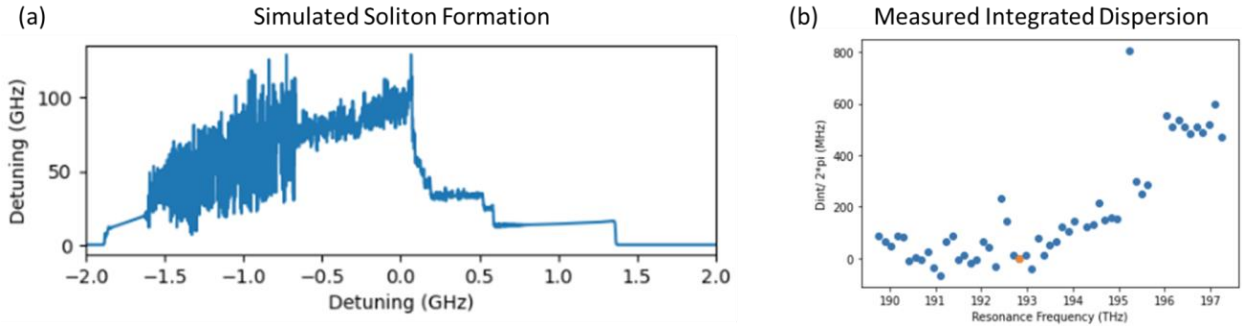

**Figure S13** (a) Simulated Intra-cavity power versus laser-resonator detuning at 600 mW pump power on chip. Multiple step-like transitions demonstrate the existence of multiple soliton states at high pump powers. (b) Integrated dispersion for the TE mode of the 2000 nm thick resonator, showing significant variance in FSR, as well as some strong avoided mode crossings.

## REFERENCES : SUPPLEMENTARY

1. Jin, W. *et al.* Deuterated silicon dioxide for heterogeneous integration of ultra-low-loss waveguides. *Opt. Lett., OL* **45**, 3340–3343 (2020).

2. UCSB Nanofabrication Facility. PECVD Recipes - UCSB Nanofab Wiki. https://wiki.nanofab.ucsb.edu/wiki/PECVD_Recipes#ICP-PECVD_.28Unaxis_VLR.29.

3. Wu, Z. *et al.* Low-noise Kerr frequency comb generation with low temperature deuterated silicon nitride waveguides. *Opt. Express, OE* **29**, 29557–29566 (2021).

4. Dergez, D., Schalko, J., Bittner, A. & Schmid, U. Fundamental properties of a-SiNx:H thin films deposited by ICP-PECVD for MEMS applications. *Applied Surface Science* **284**, 348–353 (2013).

5. Bogaerts, W. *et al.* Silicon microring resonators. *Laser & Photonics Reviews* **6**, 47–73 (2012).

6. Chrostowski, L. & Hochberg, M. *Silicon Photonics Design: From Devices to Systems*. (Cambridge University Press, 2015). doi:10.1017/CBO9781316084168.

7. Bose, D., Wang, J. & Blumenthal, D. J. 250C Process for < 2dB/m Ultra-Low Loss Silicon Nitride Integrated Photonic Waveguides. in *Conference on Lasers and Electro-Optics (2022), paper SF3O.1* SF3O.1 (Optica Publishing Group, 2022). doi:10.1364/CLEO_SI.2022.SF3O.1.

8. Puckett, M. W. *et al.* 422 Million intrinsic quality factor planar integrated all-waveguide resonator

with sub-MHz linewidth. *Nat Commun* **12**, 934 (2021).

9. Xie, Y. *et al.* Soliton frequency comb generation in CMOS-compatible silicon nitride microresonators. *Photon. Res., PRJ* **10**, 1290–1296 (2022).

10. Liu, K. *et al.* 36 Hz integral linewidth laser based on a photonic integrated 4.0 m coil resonator. *Optica, OPTICA* **9**, 770–775 (2022).

11. Brodnik, G. M. *et al.* Optically synchronized fibre links using spectrally pure chip-scale lasers. *Nat. Photon.* **15**, 588–593 (2021).

12. Chiles, J. *et al.* Deuterated silicon nitride photonic devices for broadband optical frequency comb generation. *Opt. Lett., OL* **43**, 1527–1530 (2018).

13. Chia, X. X. *et al.* Low-Power Four-Wave Mixing in Deuterated Silicon-Rich Nitride Ring Resonators. *Journal of Lightwave Technology* **41**, 3115–3130 (2023).

14. Ye, Z., Fülöp, A., Helgason, Ó. B., Andrekson, P. A. & Torres-Company, V. Low-loss high-Q silicon-rich silicon nitride microresonators for Kerr nonlinear optics. *Opt. Lett., OL* **44**, 3326–3329 (2019).

15. Gundavarapu, S. *et al.* Sub-hertz fundamental linewidth photonic integrated Brillouin laser. *Nature Photon* **13**, 60–67 (2019).

16. Moille, G., Li, Q., Lu, X. & Srinivasan, K., pyLLE: a Fast and User Friendly Lugiato-Lefever Equation Solver, *Journal of Research (NIST JRES)*, National Institute of Standards and Technology, Gaithersburg, MD, (2019), https://doi.org/10.6028/jres.124.012